 \documentclass[final,3p,times,authoryear]{elsarticle}

\usepackage{amssymb}
\usepackage{amsmath}    
\usepackage{graphicx}   
\usepackage{verbatim}   
\usepackage{color}      
\usepackage{caption}
\usepackage{subcaption}
\usepackage{hyperref}   
\usepackage{amsthm}
\usepackage{enumerate}
\usepackage{amsfonts}
\usepackage{multirow}
\usepackage{undertilde}
\usepackage{setspace}
\usepackage{appendix}

\newcommand{\bbeta}{\boldsymbol\beta}

\newcommand{\xx}{\mathbf{x}}
\newcommand{\XX}{\mathbf{X}}

\newcommand{\yy}{\mathbf{y}}

\newcommand{\g}{\mathbf{g}}

\newcommand{\s}{\mathbf{s}}

\journal{Computational Statistics and Data Analysis}

\begin{document}

\begin{frontmatter}



\title{SIMD Parallel MCMC Sampling with Applications for Big-Data Bayesian Analytics}


\author{Alireza S. Mahani}
\address{Scientific Computing Group, Sentrana Inc., Washington DC, USA \\ Phone: +1-202-507-4524}
\ead{alireza.mahani@sentrana.com}
\author{Mansour T.A. Sharabiani}
\address{Department of Epidemiology and Biostatistics, Imperial College London, UK}

\begin{abstract}
Computational intensity and sequential nature of estimation techniques for Bayesian methods in statistics and machine learning, combined with their increasing applications for big data analytics, necessitate both the identification of potential opportunities to parallelize techniques such as Monte Carlo Markov Chain (MCMC) sampling, and the development of general strategies for mapping such parallel algorithms to modern CPUs in order to elicit the performance up the compute-based and/or memory-based hardware limits. Two opportunities for Single-Instruction Multiple-Data (SIMD) parallelization of MCMC sampling for probabilistic graphical models are presented. In exchangeable models with many observations such as Bayesian Generalized Linear Models (GLMs), child-node contributions to the conditional posterior of each node can be calculated concurrently. In undirected graphs with discrete nodes, concurrent sampling of conditionally-independent nodes can be transformed into a SIMD form. High-performance libraries with multi-threading and vectorization capabilities can be readily applied to such SIMD opportunities to gain decent speedup, while a series of high-level source-code and runtime modifications provide further performance boost by reducing parallelization overhead and increasing data locality for Non-Uniform Memory Access architectures. For big-data Bayesian GLM graphs, the end-result is a routine for evaluating the conditional posterior and its gradient vector that is 5 times faster than a naive implementation using (built-in) multi-threaded Intel MKL BLAS, and reaches within the striking distance of the memory-bandwidth-induced hardware limit. Using multi-threading for cache-friendly, fine-grained parallelization can outperform coarse-grained alternatives which are often less cache-friendly, a likely scenario in modern predictive analytics workflow such as Hierarchical Bayesian GLM, variable selection, and ensemble regression and classification. The proposed optimization strategies improve the scaling of performance with number of cores and width of vector units (applicable to many-core SIMD processors such as Intel Xeon Phi and Graphic Processing Units), resulting in cost-effectiveness, energy efficiency (`green computing'), and higher speed on multi-core x86 processors.
\end{abstract}


\end{frontmatter}


\section{Introduction} \label{sec-intro}
Many inference problems in statistics and machine learning are best expressed in the language of probabilistic graphical models, where a probability distribution function (PDF) over a high-dimensional parameter space can be motivated as the product of a collection of terms, each a function of a subset of the parameters. Inference in such models requires summarizing the joint PDF, which can be quite complex and lacking closed-form integrals. Monte Carlo Markov Chain (MCMC) sampling techniques offer a practical way to summarize complex PDFs for which exact sampling algorithms are not available. For many real-world problems, MCMC sampling can be very time-consuming due to a combination of large data sets, high dimensionality of joint PDF, lack of conjugacy between likelihood and prior functions, and poor mixing of the MCMC chain. Fast MCMC sampling is, therefore, important for wider adoption of probabilistic models in real-world applications.

For many years, software developers could rely on faster processors to see improved performance, without any need for code modification. In the past decade, however, chip manufacturers have warned of single-core performance saturation as CPU clock rates and instruction-level parallelism reach their physical limits (\cite{jeffers2013intel}). Instead, we are seeing a steady rise in core counts and width of vector units, culminating in the emergence of many-core Single-Instruction-Multiple-Data (SIMD) architectures such as Graphic Processing Units (GPUs) \footnote{\url{http://www.nvidia.com/object/what-is-gpu-computing.html}} and Intel Xeon Phi \footnote{\url{http://tinyurl.com/co9e8hy}}. Today, a CPU purchased for around \$3000 can offer a theoretical peak double-precision performance of more than 300 GFLOPS \footnote{\url{http://ark.intel.com/products/64595/}}, and many-core processors selling for nearly the same price have broken the Tera FLOP barrier \footnote{\url{http://tinyurl.com/oexotqv}}. For scientific computing applications to take full advantage of such enormous performance potential within a single compute node, the entire parallelism potential of an algorithm must be efficiently exposed to the compiler, and eventually to the hardware.

Most efforts on parallelizing MCMC have focused on identifying high-level parallelism opportunities such as concurrent sampling of conditionally-independent nodes (\cite{wilkinson2010parallel}). Such coarse-grained parallelism is often mapped to a distributed-memory cluster or to multiple cores on a shared-memory node. As such, vectorization capabilities of the processor are implicitly assumed to be the responsibility of libraries and compilers, resulting in a systemic under-utilization of vectorization in scientific computing applications. Furthermore, with increasing data sizes and widening gap between floating-point performance and memory bandwidth, modern processors have seen an architectural change in memory layout from Symmetric Multi-Processors (SMPs) to Non-Uniform Memory Access (NUMA) designs to better scale total memory bandwidth with the rising core count. The software-level, shared-memory view of this asymmetric memory is a convenient programming feature but can lead to a critical memory bandwidth uder-utilization for data-intensive applications. This paper seeks to expand our ability to make efficient use of multicore x86 processors - today's most ubiquitous computing platform - to rapidly generate samples from the posterior distribution of model parameters. Our focus is two-fold: 1) identifying opportunities for SIMD parallelism in MCMC sampling of graphical models, and 2) efficiently mapping such SIMD opportunities to multi-threading and vectorization parallel modes on x86 processors. Using examples from directed and undirected graphs, we show that off-the-shelf, multi-threaded and vectorized high-performance libraries (along with vectorizing compilers) provide a decent speedup with small programming effort. Additionally, a series of high-level source-code and runtime modifications lead to significant additional speedup, even approaching hardware-induced performance limits. Vectorization of SIMD parallelism opportunities are often complementary with coarse-grained parallel modes and create a multiplicative performance boost. Moreover, we illustrate the counter-intuitive result that, given a limited number of cores available, efficient fine-grained (SIMD) multi-threading can outperform coarse-grained multi-threading over a range of data sizes where L3 cache utilization is the dominating factor. In the process, we propose a general strategy for optimally combining a sequence of maps to minimize multi-threading overhead while maximizing vectorization coverage. This strategy naturally allows for data locality at the memory and cache level, and significantly reduces the cost of complex synchronizations such as reduction on arrays. Furthermore, a differential update strategy for MCMC sampling of graphical models is proposed which, where applicable, can lead to significant reduction in data movement as well as the amount of computation, applicable to graphs with continuous as well as discrete nodes.

The remainder of this paper is organized as follows. In Section \ref{sec-pargibbs-theory} we lay the theoretical foundation for the paper, including a summary of previous research on parallel MCMC and overview of two single-chain parallelization techniques for parallel MCMC of Directed Acyclic Graphs (DAGs). In Section \ref{sec-pargibbs-glm} we do a detailed performance analysis and optimization of parallel MCMC for Bayesian GLM (focusing on logistic regression). In Section \ref{sec-ext} we discuss several extensions to the ideas developed in Section \ref{sec-pargibbs-glm}, including Hierarchical Bayesian (HB) GLM, calculation of derivatives, Ising model, batch RNG, distributed and many-core computing, and compile-time loop unrolling. Section \ref{sec-discussion} contains a summary of our results and pointers for future research.

\section{Parallel MCMC for Graphical Models} \label{sec-pargibbs-theory}
\subsection{Previous Work on Parallel MCMC} \label{subsec-litrev}
A straightforward method for parallel MCMC is to run multiple chains (\cite{wilkinson2010parallel}). Since each chain must go through the burn-in period individually, multi-chain parallelization is less suitable for complex models with poor convergence where a big fraction of time might be spent on the burn-in phase. The most common single-chain MCMC parallelization strategy is concurrent sampling of conditionally-independent nodes within the Gibbs sampling framework (\cite{wilkinson2010parallel}). This is also known as graph coloring or chromatic sampling. Example applications include Hierarchical LDA models (\cite{newman2009distributed}) and Markov Random Fields (\cite{gonzalez2011parallel}). In molecular dynamics simulations, assumption of short-range intermolecular forces provides for similar domain decompositions (\cite{ren2007parallel}). This parallelization strategy is reviewed in Section \ref{subsec-simd-gibbs}. In synchronous Gibbs sampling (\cite{geman1984stochastic}), all variables are updated simultaneously - rather than sequentially - in each Gibbs cycle, regardless of whether conditional independence holds or not. Despite its theoretical and practical shortcomings, this embarrassingly parallelizable approach remains popular in large-scale MCMC problems such as Latent Dirichlet Allocation (LDA) models for topical analysis of text documents (\cite{newman2009distributed}). Parallel message-passing algorithms are discussed in \cite{gonzalez2009distributed,gonzalez2011parallel,gonzalez2012powergraph} with application to discrete factor graphs. \cite{tibbits2011parallel} present CPU and GPU parallelization of multivariate slice sampling (\cite{neal2003slice}) with application to linear Gaussian processes. \cite{xu2011multicore} propose a parallel `look-ahead' sampler for discrete-valued densely-connected graphical models such as Boltzmann Machines and LDA. This approach takes advantage of the observation that, in densely-connected networks, contribution from any given node to the conditional posterior of a target node is small. A similar strategy, called `pre-fetching' has been proposed (\cite{brockwell2006parallel,strid2010efficient}) in the context of one-block Metropolis-Hastings algorithms. \cite{yu2009parallel} propose an approximate parallel Gibbs sampler - with GPU implementation - for DNA motif finding, where parallelization is achieved through replacing sequential random steps by a deterministic, pre-ordered search of the parameter space.

In terms of target platforms, MCMC parallelization work has primarily focused on shared-memory multi-threading (\cite{xu2011multicore,tibbits2011parallel}), distributed-memory parallelization using explicit or implicit message-passing (\cite{ren2007parallel,ahmed2012scalable,low2012distributed,gonzalez2009distributed,newman2009distributed}), and more recently, many-core architectures such as GPUs (\cite{yu2009parallel,tibbits2011parallel,da2011cudabayesreg,dumont2011markov}). This leaves two areas under-explored: efficient use of vector instructions for SIMD parallelization, and NUMA awareness for efficient utilization of memory bandwidth. Addressing these two areas is important in light of trends towards higher core counts and multi-socket processors with NUMA architecture, and increasing width of vector units. Today's desktop processors are increasingly dual-socket, while quad-socket setups are available for high-end servers. Since each socket has a separate memory controller, a processor's access to memory attached to other sockets is mediated through a slower path such as Intel QPI \footnote{\url{http://tinyurl.com/7hhc7rj}}, thus creating a non-uniform access to memory. On the other hand, vector units have steadily widened over recent generations to reach 128 bits for SSE4, and 256 bits for AVX instruction set extensions. Intel Xeon Phi co-processors have doubled the width yet again to 512 bits. The combined effect of vectorization and NUMA awareness for a dual-socket x86 can reach an order of magnitude: A fully-vectorized code can run up to 4 times faster for double-precision calculations, while minimizing cross-socket communication on a dual-socket processor can double the performance for a memory-bound application. Our paper aims to help better utilize single-server capacities of a modern x86 processor for parallel MCMC sampling. In addition to achieving more efficient execution on x86 processors, our performance optimization techniques pave the way for further speedup on many-core processors since the underlying principles of reducing parallel overhead and increase data locality apply to all architectures. This strategy has been dubbed `transforming and tuning' (\cite{jeffers2013intel}).

\subsection{Gibbs Sampling of Graphical Models} \label{subsec-overview}
A graphical model consists of \emph{nodes} (or \emph{vertices}) connected by \emph{links} (or \emph{edges}) \footnote{Our introduction to graphical models borrows heavily from Section 8.1 of Bishop's excellent textbook (\cite{christopher2006pattern}).}. In a probabilistic graphical model, each node represents a random variable (or a group of random variables), and links express probabilistic relationship between these variables. The graph represents a decomposition of the joint probability distribution over all variables into a product of factors, each consisting of a subset of variables. In DAGs, links have directionality indicated by arrows, and no closed path exists such that one can always travel in the direction of arrows. For a DAG with $K$ nodes, the joint distribution is given by the following factorization
\begin{equation} \label{eq:dag}
p(\xx) = \prod_{k=1}^K p(x_k | \mathrm{pa}_k),
\end{equation}
where $\mathrm{pa}_k$ represents the set of parents of node $x_k$, and $\xx=\{x_1,\cdots,x_K\}$. For undirected graphs, the joint distribution is expressed as a product of potential functions $\psi_C(\xx_C)$ over maximal cliques of the graph:
\begin{equation}
p(\xx) = \frac{1}{Z} \prod_C \psi_C(\xx_C),
\end{equation}
where $Z$ is the normalization factor, also known as the partition function.

Inference and prediction in graphical models often requires calculating the expected values of functions of $\xx$ with respect to the joint distribution $p(\xx)$. In many cases, closed forms for integrals do not exist and approximate techniques such as MCMC can be used to approximate the joint distribution with a finite sample. A popular MCMC technique is Gibbs sampling (\cite{geman1984stochastic}) which is a special case of the Metropolis-Hastings (MH) algorithm (\cite{hastings1970monte}) with 100\% acceptance rate. In Gibbs sampling, we iterate through the stochastic nodes and draw samples from the probability distribution for each node, conditioned on the latest samples drawn for all remaining nodes. Below is an outline of the Gibbs sampling algorithm:
\begin{enumerate}
\item Initialize all nodes $\xx=\{x_1,\cdots,x_K\}$.
\item For iteration $t=1,\cdots,T:$
\begin{itemize}
\item Sample $x_1^{t+1} \sim p(x_1 | x_2^t,x_3^t,\cdots,x_K^t)$.
\item Sample $x_2^{t+1} \sim p(x_2 | x_1^{t+1},x_3^t,\cdots,x_K^t)$.
\item $\vdots$
\item Sample $x_k^{t+1} \sim p(x_j | x_1^{t+1},\cdots,x_{k-1}^{t+1},x_{k+1}^t,\cdots,x_K^t)$.
\item $\vdots$
\item Sample $x_K^{t+1} \sim p(x_M | x_1^{t+1},x_2^{t+1},\cdots,x_{K-1}^{t+1})$.
\end{itemize}
\end{enumerate}
Gibbs sampling is appealing because it reduces the problem of sampling from a high-dimensional distribution to a series of low-dimensional sampling problems, which are often easier, especially when the resulting low-dimensional distributions have standard sampling algorithms (e.g. multivariate Gaussian, Gamma, etc.). When conditional distributions cannot be directly sampled, a hybrid approach can be used where an MCMC algorithm such as Metropolis (\cite{metropolis1949monte}) or slice sampler (\cite{neal2003slice}) is applied within the Gibbs framework (\cite{robert2004monte}).

\subsection{SIMD Parallel MCMC for DAGs} \label{subsec-simd-gibbs}
Here we present two canonical parallelization opportunities for DAGs to provide a foundation for our analysis of Bayesian GLM. Equivalent ideas for undirected graphs are discussed in \ref{subsec-beyond-glm} in the context of Ising model and Boltzmann machine.

According to Eq. \ref{eq:dag}, a given node appears in the joint distribution in two ways: a) self-term, i.e. the term that expresses the probability of the node conditioned on its parents, and b) child terms, i.e. terms corresponding to the conditional probability of the node's children. Therefore, the conditional distributions appearing in Gibbs sampling algorithm can be decomposed as follows:
\begin{equation} \label{eq:self-child-terms}
p(x_k|-) \propto \overbrace{ p(x_k|\mathrm{pa}_k)}^\text{self term} \:\: \times \overbrace{\prod_{j \:\mathrm{s.t.}\: x_k \in \mathrm{pa}_j} p(x_j | \mathrm{pa}_j)}^\text{child terms}
\end{equation}
This decomposition allows us to identify the following two single-chain parallelization modes for Gibbs sampling of DAGs.

{\parindent0pt \textit{Parallel Sampling of Conditionally-Independent Nodes}}: Consider two nodes $x_1$ and $x_2$. If their joint conditional distribution is factorizable, i.e. if $p(x_1,x_2|-) = p_1(x_1|-)p_2(x_2|-)$, then the nodes are called \emph{conditionally-independent} and can be sampled simultaneously in each iteration of Gibbs sampling. From Eq. \ref{eq:self-child-terms}, we can infer the following theorem about conditional independence:
\newtheorem{theorem:ci-parallelism}{Theorem}[section]
\begin{theorem:ci-parallelism} \label{theorem:ci-parallelism}
Two nodes are conditionally-independent (and can be sampled in parallel) if a) neither node is a parent of the other, AND b) nodes have no common children.
\end{theorem:ci-parallelism}
\begin{proof}
Condition (a) means that neither node appears in the self term of the other node. Condition (b) means that neither node appears in any of the child terms for the other nodes. Therefore, neither node appears in the conditional distribution of the other node. This is sufficient to prove that nodes are conditionally-independent and can be sampled concurrently in Gibbs sampling.
\end{proof}
The above theorem can be used to determine if any pair of nodes are conditionally-independent or not. However, constructing the most efficient partitioning of a graph into node blocks, each of which consists of conditionally-independent nodes is non-trivial and an NP-hard problem (\cite{wilkinson2010parallel}). For specific graph structures, such as HB models (\cite{peter2005bayesian}), conditionally-independent node blocks can be constructed rather easily. This parallelization mode can be difficult to vectorize, especially for non-conjugate, complex posteriors (such as GLM) where drawing a sample for each node involves multiple function calls including random number generation (RNG) and a sampling routine. As we see for Ising models, however, for simpler distributions, this parallel mode can be partially vectorized using the strategy discussed in Section \ref{subsec-perfopt}, with significant contribution towards overall performance of the sampling routine. 

{\parindent0pt \textit{Parallel Calculation of Conditional Posterior}}: First, consider the conjugate case, where the product of prior and each child-term contribution retains the functional form of the prior. For example, in LDA models a Dirichlet prior is used on the distribution of topics in each document. The likelihood contribution of each token in a document is a categorical distribution, with their collective contribution being a multinomial distribution whose parameter vector is the sum of contributions from all tokens. Similarly, with a conjugate Dirichlet prior on distribution of words in each topic, the contribution from each document (and tokens within each document) takes an additive form towards the posterior. In such cases, calculating the parameters of the posterior distribution can be parallelized, followed by a reduction. Second, consider the case where conjugacy does not exist, and rather than drawing independent samples from the posterior we must create a Markov chain that converges to the posterior. For most MCMC sampling techniques such as Slice sampler (\cite{neal2003slice}) or Metropolis sampler (\cite{metropolis1949monte}), the majority of time during each sampling cycle is spent on evaluating the conditional log-posterior function (e.g. within the Gibbs sampling framework) or its derivatives (Section \ref{subsec-deriv}). Therefore, any speedup in function evaluation translates, nearly 1-to-1, into MCMC speedup. Similar to the conjugate case, the additive contribution of child terms can be calculated in parallel, followed by a reduction step. In both cases, when the model is exchangeable (\cite{bernardo1996concept}) as in the case of i.i.d. observations, the contributions from child nodes are symmetric, resulting in a SIMD computation.

Non-conjugate cases are typically better candidates for SIMD parallelization since 1) they tend to take a larger fraction of each sampling cycle, 2) in conjugate cases, we need  a larger number of observations before the cost of calculating the conjugate posterior parameters dominates the cost of generating a random deviate from the posterior. GLM models are good candidates for SIMD Gibbs sampling due to non-conjugacy of prior and likelihood functions, and due to high computational cost of each function evaluation, especially for long and wide data and in the presence of transcendental functions (Section \ref{subsec-setup}).

\begin{center}
*********
\end{center} 

\begin{figure}
\centering
\begin{subfigure}[b]{3.5in}
\centering
\includegraphics[scale=0.35]{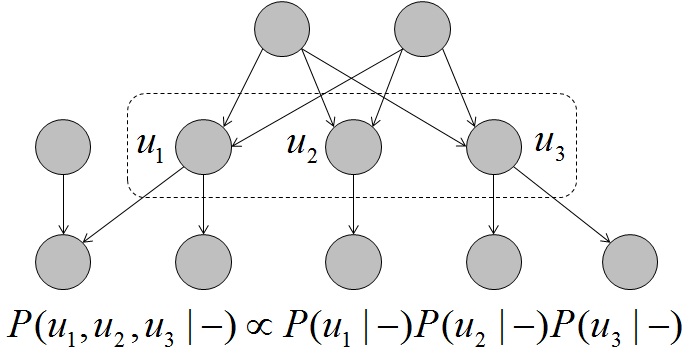}
\caption{Parallel sampling of conditionally-independent nodes: Such nodes cannot have common children, or have parent-child relationship amongst themselves.}
\label{fig:pargibbs-ci}
\end{subfigure}
\begin{subfigure}[b]{3.5in}
\centering
\includegraphics[scale=0.35]{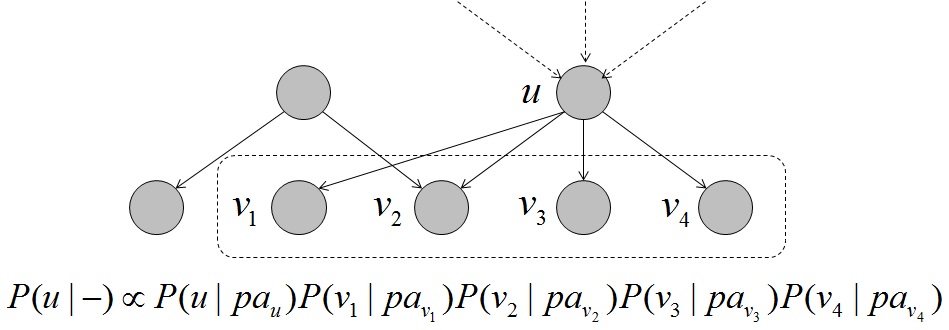}
\caption{Parallel computation of conditional posterior: Contribution from child nodes can be calculated concurrently. If parent-child relationships are symmetric, this parallelization mode takes a SIMD form.}
\label{fig:pargibbs-simd}
\end{subfigure}
\caption{Two parallelization modes for MCMC sampling of probabilistic DAGs.}
\label{fig:parallel-gibbs}
\end{figure}

Figure \ref{fig:parallel-gibbs} summarizes the two single-chain parallelization modes for DAGs discussed above. In relative terms, parallel sampling of conditionally-independent nodes can be considered coarse-grained, while parallel evaluation of log-likelihood can be considered fine-grained. In the remainder of this paper, we use two examples - Bayesian GLM and Ising model - to demonstrate how each of these modes can be parallelized on multi-core x86 processors using multi-threading and vectorization.

\section{SIMD Parallel MCMC for Bayesian GLM on Multi-core x86} \label{sec-pargibbs-glm}
\subsection{Setup} \label{subsec-setup}
\subsubsection{Running Example: Bayesian Logistic Regression} \label{subsubsec-setup-glm}
GLMs (\cite{nelder1972generalized}) are the workhorse of statistics and machine learning, with applications such as risk analysis (\cite{sobehart2000moody,fenton2004combining,antonio2007actuarial}) and public health (\cite{hastie1987non,azar2011immunologic}) among others. They can be extended to handle data sparseness and heterogeneity via HB framework (\cite{peter2005bayesian,gelman2007data}), or to account for repeated measurements and longitudinal data via Generalized Linear Mixed Models (\cite{mcculloch2006generalized}).

\begin{figure}
\begin{centering}
\includegraphics[scale=0.65]{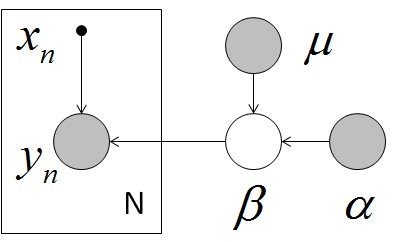}
\caption[]{DAG representing the logistic regression problem. We use the plate notation to describe an array of nodes.}
\label{fig:logit-dag}
\end{centering}
\end{figure}

For our purposes, GLM models have the following log-likelihood function:

\begin{equation}
L(\bbeta) = \sum_{n=1}^N f(\xx_n^t \bbeta; y_n),
\end{equation}
where $\xx_n$ is the vector of features or explanatory variables for observation $n$ and has a length $K$, $y_n$ is the response or target variable, $\bbeta$ is the coefficient vector of length $K$, and $N$ is the number of observation points. Logistic regression is an important member of GLM family, where each binary response variable $y_i$ is assumed to follow a Bernoulli distribution whose probability is $1/(1+\exp(-\xx_n^t \bbeta))$. The log-likelihood becomes
\begin{equation} \label{eq-loglike-binlogit}
L(\bbeta) = - \sum_{n=1}^N \left\{ (1-y_n) \xx_n^t \bbeta + \log(1+\exp(-\xx_n^t \bbeta)) \right\},
\end{equation}
with $y_n$ being a binary response variable with possible values $\{0,1\}$. In a Bayesian setting, we impose a prior distribution on $\bbeta$ such as a multivariate Gaussian with parameters $\alpha$ and $\mu$, leading to the DAG representation of Fig. \ref{fig:logit-dag}. Unless we are dealing with very small data sizes, calculation of prior contribution is often computationally negligible compared to the likelihood calculation, and hence we focus on the latter throughout this paper.

\subsubsection{Hardware} \label{subsubsec-setup-hardware}
We use a dual-socket 2.6GHz Intel Xeon E5-2670 \footnote{\url{http://ark.intel.com/products/64595/}} with 8 cores per socket (16 total cores), 20MB L3 cache per processor (shared among 8 cores), 256KB of L2 cache per core, and 32KB of L1 cache per core (separately for data and instructions each). Total installed RAM is 128GB, with maximum memory bandwidth of 51.2GB/sec per socket. Since each processor has its own memory controller, the dual-socket setup creates a Non-Uniform Memory Access (NUMA) environment (Section 8.8.2 of Intel Optimization Reference Manual \footnote{\url{http://tinyurl.com/7ty2lsx}}). This processor supports the AVX instruction set extensions, using 256-bit wide floating-point registers YMM0-YMM15. Hyper-threading is turned off for our experiments.

\subsubsection{Software} \label{subsubsec-setup-software}
All code is written in C/C++ \footnote{During performance measurement, all BLAS calls are made to FORTRAN routines. For presentation brevity, however, we show calls to the C interface of BLAS in this paper.}. (The only C++ feature used is template metaprogramming for compile-time loop unroll. See Section \ref{subsubsec-jit-unroll}.) We use the Intel software stack for compiling and micro-benchmarking as well as high-performance mathematical routines:
\begin{itemize}
\item Intel Math Kernel Library (MKL) (part of Intel Composer XE 2013): We use MKL for vectorized and multithreaded BLAS calls (dgemv/ddot), vectorized transcendental functions (as part of Vector Math Library or VML \footnote{\url{http://tinyurl.com/nc9yjhh}}), and vectorized random number generation (RNG) as part of Vector Statistical Library or VSL \footnote{\url{http://tinyurl.com/pevz69r}}. Switching between single-threaded and multi-threaded MKL is done via the link option '-mkl=sequential' or '-mkl=parallel'.\footnote{For more on how to compile and link using Intel C++ compiler, see \url{http://tinyurl.com/lctboe3}.}
\item Intel C++ Compiler (also part of Intel Composer XE 2013): Combined with Intel VML, the compiler allows us to vectorize functions of transcendental functions. All codes tested in this paper were compiled using optimization flag \texttt{-O2}.
\item Intel Performance Counter Monitor (PCM)\footnote{\url{http://tinyurl.com/q4er8qp}}, an open-source C++ wrapper library for reading micro-benchmarking (core and uncore) events from Intel's Performance Monitoring Units, including cache hit rate, clock cycles lost to memory stalls, and instructions per cycle.
\end{itemize}

\subsubsection{Parallel Programming} \label{subsubsec-setup-parallel}
Our overall approach is to maintain software portability and developer productivity while achieving good performance. For multi-threading, we use the OpenMP API and Intel's implementation of it (as part of Intel Composer XE 2013). OpenMP offers a mix of high-level and low-level compiler directives and runtime functions for creating and managing threads, and offers an efficient alternative to working directly with Pthreads. OpenMP is perhaps today's most-widely used API for multi-threading in HPC applications. Other notable options for multi-threading include Cilk Plus \footnote{\url{http://www.cilkplus.org/}} (also handles vectorization) and Threading Building Blocks \footnote{\url{https://www.threadingbuildingblocks.org/}} from Intel, Chapel \footnote{\url{http://chapel.cray.com/}} from Cray, X10 \footnote{\url{http://x10-lang.org/}} from IBM, and Phoenix MapReduce \footnote{\url{http://mapreduce.stanford.edu/}} from Stanford. For vectorization, we choose `guided auto-vectorization' \footnote{\url{http://tinyurl.com/qbtpc38}}, i.e. we use pragmas and code modifications to guide the compiler to vectorize the code. We find this option to strike a balance between performance and portability/productivity. Other options include vector intrinsics and assembly code. Another parallelization option is distributed-computing on a server cluster using Message-Passing Interface (MPI) \footnote{\url{http://www.mcs.anl.gov/research/projects/mpi/}}, MapReduce \footnote{\url{http://research.google.com/archive/mapreduce.html}} or Hadoop \footnote{\url{http://hadoop.apache.org/}}. Given the fine-grained nature of our SIMD parallelization approach and significant network latency and bandwidth restrictions in distributed computing, however, focusing within a single server is the logical choice. Our parallelization concepts, however, can be combined with coarse-grained options and thus incorporated into distributed computing frameworks. See section \ref{subsec-manycore}.

Note that while we are using Intel's hardware and software, the optimization lessons learned are applicable to other platforms, including open-source software. More importantly, the software itself is portable since we are using standard APIs (C/C++ standard library, OpenMP, BLAS).

\subsection{Baseline Implementation} \label{subsec-baseline}
To calculate the log-likelihood in Eq. \ref{eq-loglike-binlogit}, we can consolidate $\xx_n$'s into rows of a $N \times K$ matrix $\XX$, and use a matrix-vector multiplication routine such as BLAS \texttt{dgemv} to calculate $\XX \bbeta$. We refer to this first map operation (\cite{mccool2012structured}) as the Linear Algebra (LA) map. This is followed by an element-wise transformation of $\XX \bbeta$ involving transcendental functions, and a final reduction step. We call this second map operation the Transcendental (TR) map. The result is the \texttt{loglike} routine in Fig. \ref{fig:code-baseline}. Since each map operation is finished before the start of the new map, this strategy is a Sequence of Maps (SOM) pattern (\cite{mccool2012structured}). Using an off-the-shelf, optimized BLAS library such as Intel MKL provides automatic vectorization and multithreading of the \texttt{dgemv} routine. Further parallelization can be achieved by multithreading the TR map using the \texttt{omp parallel for} pragma. This same loop can also be vectorized, but it requires access to a vector math library for transcendental functions, as well as compiler support for vectorization of functions of transcendental functions. The Intel Composer suite offers both such capabilities. The \texttt{simd} pragma in Fig. \ref{fig:code-baseline} directs the Intel C++ compiler to ignore any potential vectorization barriers (such as pointer aliasing between \texttt{Xbeta} and \texttt{y}) and generate vector instructions for the TR map. \footnote{SIMD constructs have been incorporated into OpenMP 4.0 API: \url{http://www.openmp.org/mp-documents/OpenMP4.0.0.pdf}}.

\begin{figure}
\begin{verbatim}
--------------------------------------------------------------------------
double loglike(double beta[], double X[], double y[], int N, int K) {
  double r=0.0;
  double *Xbeta = new double[N];
  
  /* Linear Algebra (LA) Map */
  cblas_dgemv(CblasRowMajor,CblasNoTrans,N,K,1.0,X,K,beta,1,0.0,Xbeta,1);  
  
  /* Transcendental Loop (TR) Map + Reduction */
  #pragma omp parallel for reduction(+:r) num_threads(NTHD) schedule(static)
  #pragma simd reduction(+:r)
  for (int n=0; n<N; n++) {
    r -= (log(1.0+exp(-Xbeta[n]))+(1.0-y[n])*Xbeta[n]);
  }

  delete [] Xbeta;
  return r;
}
--------------------------------------------------------------------------
\end{verbatim}
\caption[]{Baseline (SOM) implementation of the log-likelihood function for the Bayesian logistic regression problem, using the level-2 BLAS call \texttt{dgemv} for matrix-vector multiplication. It is assumed that the \texttt{K} covariates for observation \texttt{n} are stored in \texttt{K*sizeof(double)} bytes of memory starting at address \texttt{X+n*K}. This means a row-major storage which performs slightly better in the \texttt{dgemv} call. Parameter `NTHD' is passed as a global or environment variable.}
\label{fig:code-baseline}
\end{figure}

\begin{figure}
\begin{centering}
\includegraphics[scale=0.65]{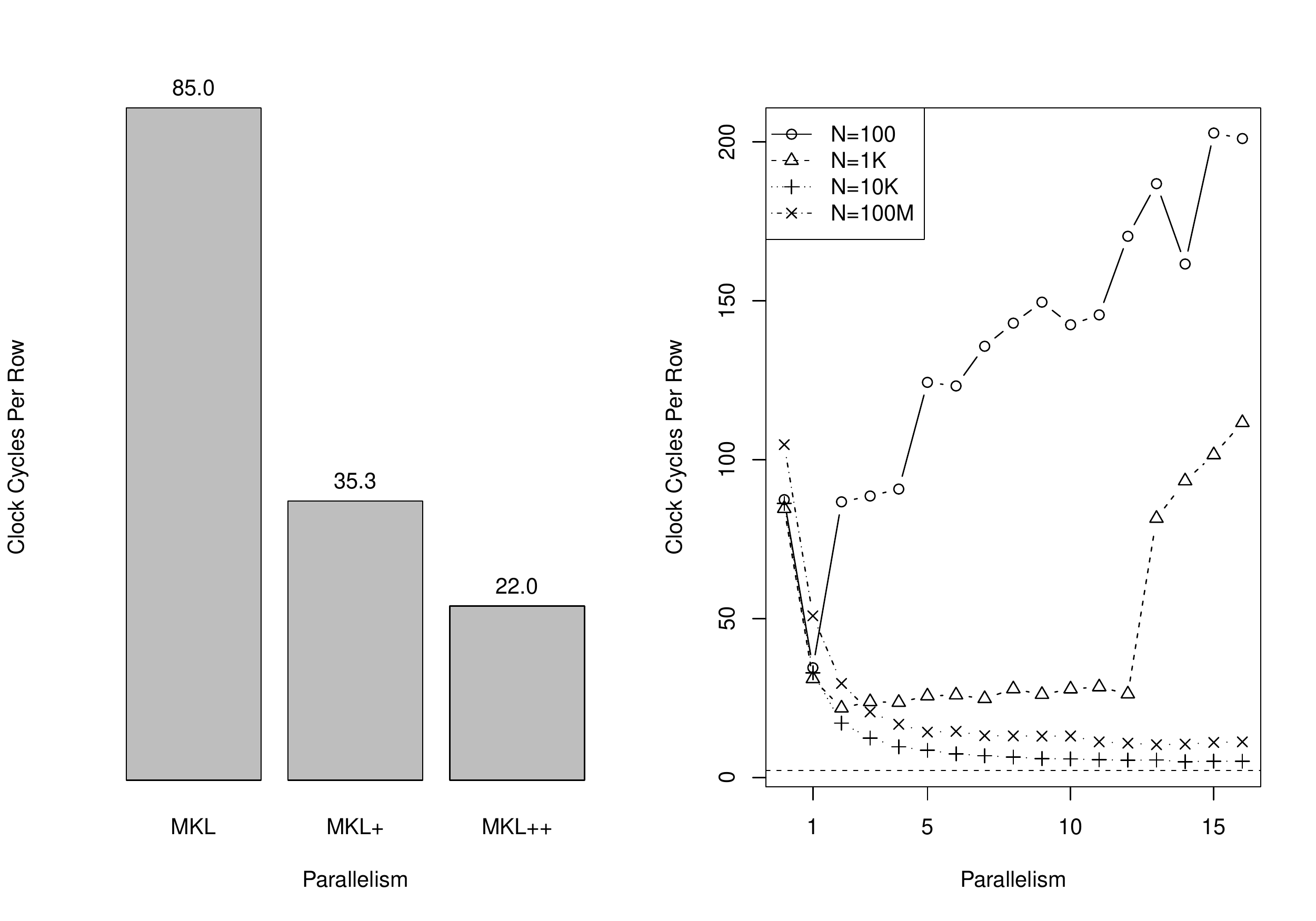}
\caption[]{Left Panel: Performance (measured in Clock Cycles Per Row or CPR) of baseline implementation for evaluating log-likelihood function for logistic regression with parameters $N=1000, K=10$. In `MKL', we use Intel MKL-BLAS to fully parallelize the LA map. In `MKL+', the TR map has been multi-threaded with 4 threads using \texttt{omp parallel for} pragma. In 'MKL++', we have used Intel Compiler and Vector Math Library to further parallelize TR through AVX vectorization, using \texttt{simd} pragma. See Fig. \ref{fig:code-baseline} for details. Right Panel: CPR as a function of parallelism, for 4 values of $N$ and for $K=10$. Parallelism is defined as follows: 0 means no TR parallelization; 1 means TR is vectorized; 2-16 indicate number of threads used in multi-threading of TR, while keeping it vectorized. Dotted horizontal line represents the memory-based lower-bound on CPR for large data. See Section \ref{subsubsec-perfanal-limits} for derivation.}
\label{fig:blas-cpr}
\end{centering}
\end{figure}

Fig. \ref{fig:blas-cpr} (left) shows the performance of our baseline implementation, measured in `Clock Cycles Per Row' (CPR), defined as total time (measured in CPU clock cycles) for each function evaluation, divided by $N$, the number of rows in $\XX$ \footnote{This is inspired by the Clock Cycles Per Element (CPE) metric, defined in \cite{bryant2003computer}, Chapter 5.}. Time per function evaluation is an average over many iterations. All iterations share the same $\XX$ and $\yy$ - allowing for caching - but $\bbeta$ is refreshed from one iteration to the next. This mimics the real-world application in MCMC sampling, where observed data does not change throughout the sampling iterations but the coefficients are refreshed. We see that simply adding the OpenMP pragma to the TR loop leads to a 2.4x speedup (MKL+), while adding the \texttt{simd} pragma to the TR loop provides another 1.6x speedup (MKL++), for a total of 3.9x compared to the automatic parallelization of LA offered by Intel MKL library.

Achieving 4x speedup by adding two lines of code (and spending a few hundred dollars on the Intel C++ compiler) is a great return on investment. But we also see two alarming trends, depicted in Fig. \ref{fig:blas-cpr} (right): 1) For small data sizes ($N=100,1000$), performance scales poorly as we increase number of threads, 2) As data sizes increase towards $N=100M$, absolute performance  decays. It appears that we have both a small-data and a big-data problem. Next we do a deeper performance analysis of the code, paving the way for optimization strategies discussed in Section \ref{subsec-perfopt}. These optimizations will add significant speedup to the baseline performance achieved by the code in Fig. \ref{fig:code-baseline}.

\subsection{Performance Analysis} \label{subsec-perfanalysis}
In LA map, $K+1$ doubles per row are consumed \footnote{We can read $\yy$ as integer since it is binary, but for simplicity we treat it as double with negligible impact on calculations} and 1 double per row is produced. LA involves $O(K)$ multiply/add calculations per row. The TR map consumes 1 double per row and produces a scalar after reduction, but involves 2 transcendental functions (exp and log) and a few arithmetic operations; as such, it is expensive but $O(1)$, i.e. it does not scale with $K$. For small $K$'s (around 10), the TR step dominates computation but when $K$ approaches 100, the scale tips towards the LA map.

\subsubsection{Hardware-Induced Performance Limits for Big Data} \label{subsubsec-perfanal-limits}
We can use hardware specifications \footnote{\url{http://tinyurl.com/p47ytt3}} of our server to derive memory-based and compute-based upper bounds on performance of our code in the big-data regime. This regime is formally defined as combinations of $N$ and $K$ such that 1) $\XX$ is much bigger than L3 cache and must therefore be fetched entirely from memory in each iteration, 2) $K$ is so large that the LA step dominates computation, making $2K$ FLOPS ($K$ multiplications and $K$ additions) a reasonable approximation of computation performed per row. The compute-based upper-bound on performance (i.e. minimum CPR) is derived as follows:

\begin{align}
\mathrm{ComputeBasedMinCPR} &= \frac{\mathrm{FlopsPerRow}}{\mathrm{MaxFlopsPerClock}} \\
&\approx \frac{2K}{2 \times \frac{\mathrm{VectorUnitWidth(Bits)}}{64} \times \mathrm{CoresPerSocket} \times \mathrm{NumberOfSockets}} \\
&\approx \frac{2K}{2 \times 4 \times 8 \times 2} \\
&\approx \frac{K}{64}.
\end{align}
Similarly, we calculate the memory-based performance ceiling:
\begin{align}
\mathrm{MemoryBasedMinCPR} &= \frac{\mathrm{CPUClockRate} \times \mathrm{BytesPerElement}}{\mathrm{MaxMemoryBandwidth}} \\
&= \frac{\mathrm{CPUClockRate} \times 8(K+1)}{\mathrm{BytesPerChannel} \times \mathrm{ChannelsPerSocket} \times \mathrm{NumberOfSockets} \times \mathrm{MemoryClockRate}} \\
&= \frac{2.6GHz \times 8(K+1)}{8 \times 4 \times 2 \times 1.6GHz} \\
&\approx \frac{K}{5}.
\end{align}
Comparing the two bounds, we see that in the big-data limit we are memory-bound nearly by a factor of 10. Achieving this memory-based performance ceiling is non-trivial and requires code modification. We now discuss two key factors that adversely affect performance scaling of our baseline (SOM) code shown in Fig. \ref{fig:code-baseline}.

\subsubsection{Parallelization Overhead} \label{subsubsec-perfanal-paroverhd}
\begin{figure}
\begin{centering}
\includegraphics[scale=0.65]{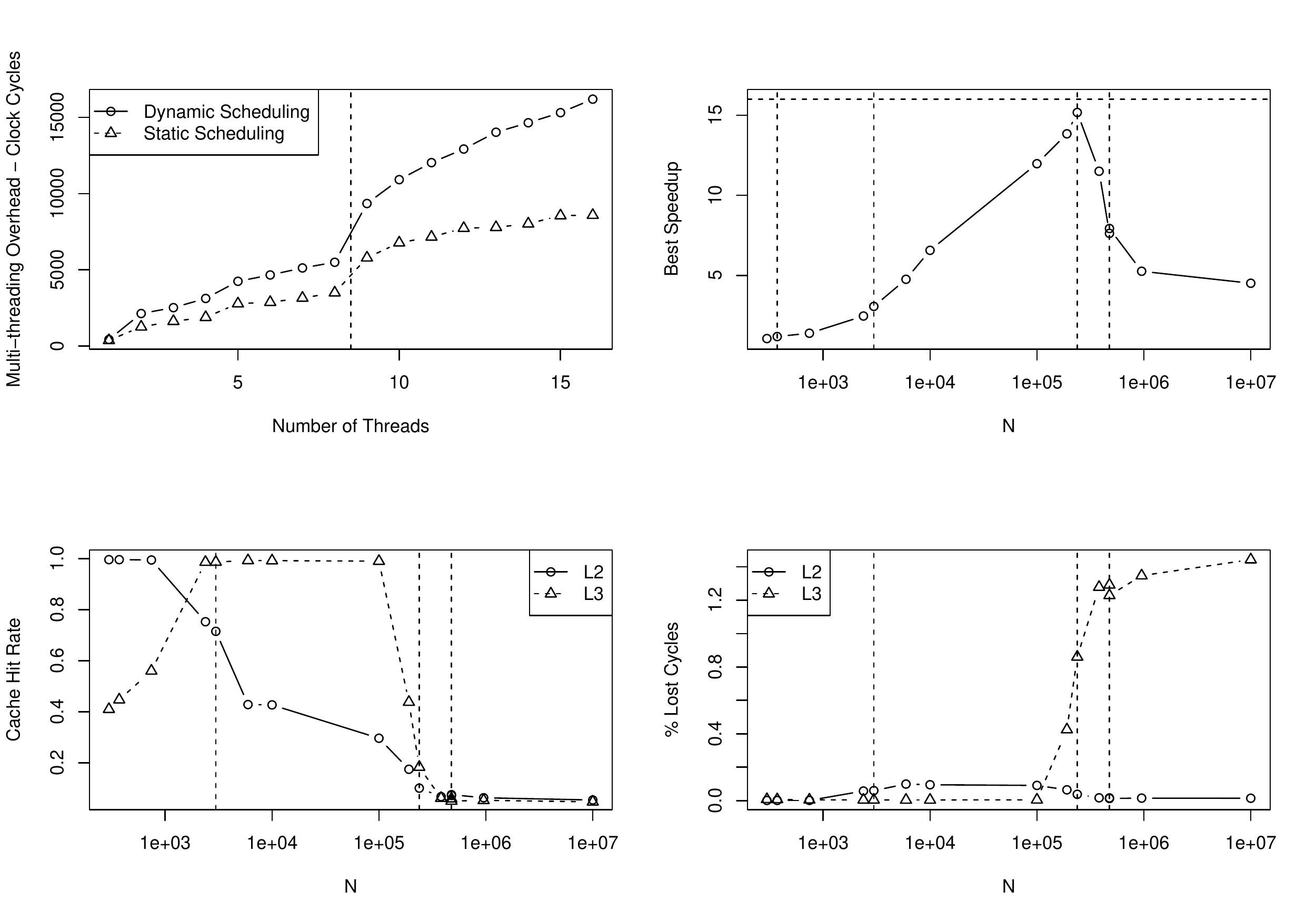}
\caption[]{Top Left: Thread management overhead, measured in CPU clock cycles, as a function of thread count, for static and dynamic scheduling. To minimize data spread, we used a compact affinity policy to generate data. Vertical line indicates transition point, to the right of which threads reside on both sockets of our dual-socket processor. Top Right: Best speedup achieved from multi-threading for $K=10$, as a function of number of observations ($N$). Horizontal line indicates the ideal outcome for a 16-core processor. Vertical lines represent $N$'s at which total size of $\XX$ and $\yy$ exceeds L1/L2/L3 cache size. Both per socket and total L3 lines are shown (two right-most lines). Bottom Panels: L2/L3 cache hit rates (left) and percent of CPU clock cycles lost due to memory stall (right) as a function of $N$ for L2/L3. Vertical lines have same interpretation as top right panel. For bottom panels, single-threaded code was used.}
\label{fig:perfanalysis-1}
\end{centering}
\end{figure}

Vectorization produces a relatively consistent improvement across the range of $N$'s plotted in Fig. \ref{fig:code-baseline} (right). Multi-threading, on the other hand, scales poorly for small $N$'s. This is expected since, in the absence of vectorization barriers such as conditional code and non-unit-stride memory access, vectorization overhead is small. In Fig. \ref{fig:perfanalysis-1} (top left) we have plotted multi-threading overhead as a function of thread count. While a constant overhead is sufficient to create an asymptotic upper bound on performance as we increase parallelism (Amdahl's law), an overhead that increases with thread count further accentuates this performance ceiling. Multi-threading overhead is amortized over $N$ to produce an equivalent CPR overhead. For example, a 10,000-clock-cycle overhead translates into a CPR overhead of 100 for $N=100$, dominating any parallelization gain for small $K$'s where the amount of computation per row is small. Larger $K$ means more computation per row and hence a smaller relative impact from multi-threading overhead. We can therefore characterize multi-threading overhead as a `small-data' problem. However, when embedded in larger frameworks such as HB, single-regression data becomes a subset of the larger data, and good performance scaling of SIMD parallel Gibbs for small to mid-size data can enhance overall performance for a big-data problem. This is discussed in detail in Section \ref{subsec-simd-embed}. We also see that multi-threading overhead is larger for dynamic scheduling compared to static scheduling. In static scheduling, jobs are pre-assigned to threads before entering parallel region, while in dynamic scheduling threads receive new job assignments during runtime as they finish previous work. Dynamic scheduling imposes more overhead due to increased need for synchronization among threads, but can be beneficial if jobs have uneven/unpredictable durations. In SIMD parallelism, job durations are often even and hence we prefer static scheduling, which is enabled by default in C/C++ specification of OpenMP, but made explicit in the code of Fig. \ref{fig:code-baseline}. Thread scheduling policy also has data locality implications, particularly for NUMA systems; this is discussed in the next section.

\subsubsection{Memory Stalls} \label{subsubsec-perfanal-memstall}
As $N$ gets larger, total size of $\XX$ and $\yy$ exceeds memory size at various levels of the hierarchy. As shown in Fig. \ref{fig:perfanalysis-1} (top right), the L3 transition marks a drastic performance degradation. Micro-benchmarking (Fig. \ref{fig:perfanalysis-1}, bottom panels) verifies that this transition phase is associated with a drop in L3 cache hit rate, and an increase in clock cycles lost to L3 cache misses. This second performance barrier can be viewed as a 'big-data' problem. Note that L2 transition has little impact on performance, as seen in speedup and lost-CPU-cycle measures. This suggests that there is sufficient computation per row to hide L2 latency. Furthermore, for data sizes that are small enough to fit into L2 the thread management overhead tends to be the dominating factor. Given these observations, we must focus on improving L3 cache utilization as well as memory bandwidth utilization. We note that there is indeed room for improving memory-bandwidth utilization, given the gap between the actual and best big-data performance, seen in Fig. \ref{fig:blas-cpr} (horizontal line in right panel). This inefficiency is inherited from the MKL-BLAS routine which is responsible for the LA map step in the log-likelihood function. This makes sense because LA map is the primary consumer of data and hence the performance bottleneck in big-data regime. We validated this claim by running \texttt{dgemv} routine alone and measuring the resulting CPR, seeing that the gap with memory-based performance ceiling remains virtually unchanged (data not shown).


\subsection{Performance Optimization} \label{subsec-perfopt}
\subsubsection{Reducing Parallelization Overhead} \label{subsubsec-perfopt-paroverhd}
The baseline code in Fig. \ref{fig:code-baseline} contains two disjoint parallel regions, one inside the \texttt{dgemv} call of LA map (with an implicit nested loop over rows and columns of $\XX$) and one in the TR map. As a result, we incur the multi-threading overhead twice, and fusing these two loops should reduce such overhead by half. A straightforward way to fuse the two loops is to move the \texttt{dgemv} call inside the TR loop by converting it to a \texttt{ddot} BLAS level 1 call. Each iteration of the fused loop serves to calculate the contribution from one row of the data towards the log-likelihood function. This routine can be considered a literal implementation of Eq. \ref{eq-loglike-binlogit}, and the outcome shown in Fig. \ref{fig:code-expanded} is an example of code/loop fusion technique used to transform a SOM into a Map of Sequences (MOS) (\cite{mccool2012structured}). While this version does reduce the multi-threading overhead, yet it can prevent full vectorization of the loop by injecting a function call inside of it. Although it is possible to improve performance through inlining and unrolling the \texttt{ddot} loop (see Section \ref{subsubsec-jit-unroll}), the MOS pattern is too inflexible for other cases such as calculating function derivatives \ref{subsec-deriv} or the Ising model \ref{subsec-beyond-glm}. We thus opt for an alternative, more general approach that reduces the multi-threading overhead of the SOM code in Fig. \ref{fig:code-baseline} while maximizing vectorization coverage.


\begin{figure}
\begin{verbatim}
--------------------------------------------------------------------------
double loglike(double beta[], double X[], double y[], int N, int K) {
  double r=0.0;
  double xbeta;
  #pragma omp parallel for reduction(+:r) num_threads(NTHD) schedule(static)
  #pragma simd reduction(+:r)
  for (int n=0; n<N; n++) {
    xbeta = cblas_ddot(K, X+n*K, 1, beta, 1);
    r -= (log(1.0+exp(-xbeta))+(1.0-y[n])*xbeta);
  }
  return r;
}
--------------------------------------------------------------------------
\end{verbatim}
\caption[]{Applying loop fusion to SOM of Fig. \ref{fig:code-baseline}, resulting in a MOS. The presence of \texttt{ddot} function call inside the for loop makes vectorization impossible or inefficient.}
\label{fig:code-expanded}
\end{figure}

The solution is presented in Fig. \ref{fig:code-hybrid}, and we refer to it as Partial Loop Fusion (PLF). It works by introducing an outer OpenMP parallel region. Each thread within the parallel region receives \texttt{N/NTHD} rows of data to process. Inside the parallel region, we switch to the single-threaded version of MKL, which remains vectorized. The TR map is explicitly vectorized (but requires no OpenMP parallelization since it is wrapped in a parallel region). We now have a single parallel region, incurring half the parallelization cost of code in Fig. \ref{fig:code-baseline}, while maintaining full vectorization. For illustrative simplicity, we have assumed that \texttt{N} is divisible by \texttt{NTHD}, but generalizations can be easily handled with a few more lines of code. The PLF approach is a general strategy for combined multi-threading and vectorization of a sequence of maps, and its applications and merits are further discussed in later sections (see Sections \ref{subsec-deriv} and \ref{subsec-beyond-glm}).

\begin{figure}
\begin{verbatim}
--------------------------------------------------------------------------
double loglike(double beta[], double X[], double y[], int N, int K) {
  double f=0.0;
  int Ny=N/NTHD, Nx=Ny*K;
  #pragma omp parallel num_threads(NTHD) reduction(+:r)
  {
    int m=omp_get_thread_num();
    double *Xbeta = new double[Ny];
    cblas_dgemv(CblasRowMajor,CblasNoTrans,Ny,K,1.0,X+m*Nx,K,beta,1,0.0,Xbeta,1);

    double rBuff=0.0;
    #pragma simd reduction(+:rBuff)
    for (int n=0; n<Ny; n++) {
      rBuff += loglike_base_f(Xbeta[n], y[m*Ny+n]);
    }
    r += rBuff;
    delete [] Xbeta;
  }
  return r;
}
--------------------------------------------------------------------------
\end{verbatim}
\caption[]{Implementation of Partial Loop Fusion (PLF) strategy for log-likelihood function of logistic regression. Within the parallel region, single-threaded but vectorized \texttt{dgemv} calls handle distinct subsets of \texttt{X} and \texttt{y}. The TR loop is explicitly vectorized. Each thread allocates private heap memory for \texttt{Xbeta}. We have assumed that \texttt{N} is a multiple of \texttt{NTHD} for code brevity.}
\label{fig:code-hybrid}
\end{figure}

\subsubsection{Reducing Memory Stall} \label{subsubsec-perfopt-memstall}
When total data size exceeds L3 cache size, subsets of data that fit in L3 must be moved in and out from memory. Therefore, in big-data regime, efficient memory bandwidth utilization is crucial. In multi-socket servers, each physical processor provides its own memory controller and cross-socket memory transactions are mediated through a slower point-to-point interconnect such as Intel QPI \footnote{\url{http://tinyurl.com/7hhc7rj}} or AMD's HyperTransport \footnote{\url{http://tinyurl.com/qxlzd4g}}. This creates asymmetric memory access latency and bandwidth depending on whether memory transaction happens within-socket or cross-socket. In contrast to Symmetric Multi-Processor Systems (SMPs) that suffer from memory bandwidth contention, NUMA topologies offer total memory bandwidth that scales with number of processors, but to utilize this total bandwidth we must localize data access within each socket as much as possible. To this end, we seek to `specialize' each socket on one half of data (assuming all 16 cores are being used), by taking the following two steps:
\begin{enumerate}
\item Split $\XX$ and $\yy$ in half, and allocate each half on the memory banks associated with one processor. The implementation is described below.
\item Utilize a compact thread-core affinity policy, which packs threads 0-7 on the first socket, and threads 8-15 on the second processor. This is done through the following system call:

\texttt{export KMP\_AFFINITY="granularity=fine,compact"}.
\item Modify the multi-threaded log-likelihood routine to point all threads on each socket to their respective socket-local arrays.
\end{enumerate}
A note on NUMA-aware memory allocation is warranted here. Commodity servers use a virtual memory system to hide the physical RAM address space and present a single linear address space to all applications. This makes memory affinity management (step 1) highly OS-specific. Providing automatic and efficient NUMA-aware memory allocation and task scheduling to applications running on multi-socket processors is an active area of research for operating systems, compilers and libraries (\cite{diaconescu2007approach,broquedis2009dynamic,ribeiro2010memory,majo2011memory,molina2011using}). The Linux header file \texttt{numa.h} \footnote{\url{http://man7.org/linux/man-pages/man7/numa.7.html}} along with the \texttt{libnuma} library offers a collection of functions for NUMA-aware memory allocation and thread scheduling. In this paper, we take advantage of the first-touch policy adopted by the Linux kernel \footnote{\url{https://software.intel.com/en-us/articles/memory-allocation-and-first-touch}}, where physical pages are allocated to the local memory controller of the thread that first 'touches' the page, i.e. reads from or writes to a memory address within that page. More specifically, we spawn 2 threads, attach each to one processor, and then use a standard heap management library such as \texttt{malloc} or \texttt{new} to allocate memory for one half of the data and copy the data from the master copy into socket-local arrays. The resulting memory allocation and log-likelihood code is shown in Fig. \ref{fig:code-numa-malloc}.

\begin{figure}
\begin{verbatim}
--------------------------------------------------------------------------
/* NUMA-aware memory allocation */  
double *X1, *X2, *y1, *y2;
int N2=N/2, NK2=N*K/2;
#pragma omp parallel num_threads(Ncore)
{
  int tid = omp_get_thread_num();
  if (tid==0) { // socket 1 gets first half of data
    X1 = new double[NK2];
    y1 = new double[N2];
    memcpy(X1, X, (NK2)*sizeof(double));
    memcpy(y1, y, (N2)*sizeof(double));
  } else if (tid==Ncore/2) { // socket 2 gets second half of data
    X2 = new double[NK2];
    y2 = new double[N2];
    memcpy(X2, X+NK2, (NK2)*sizeof(double));
    memcpy(y2, y+N2, (N2)*sizeof(double));
  }
}
...
/* NUMA-aware log-likelihood function */
double loglike(double beta[], double X1[], double X2[], double y1[], double y2[], int N, int K) {
  double f=0.0;
  int Ny=N/NTHD, Nx=Ny*K, NTHD2=NTHD/2;
  #pragma omp parallel num_threads(NTHD) reduction(+:f)
  {
    int m=omp_get_thread_num(), meff;
    double *Xp, *yp;
    if (m<NTHD2) {
      Xp=X1; yp=y1;
      meff=m;
    } else {
      Xp=X2; yp=y2;
      meff=m-NTHD2;
    }
    double *Xbeta = new double[Ny];
    double *Xtmp=Xp+meff*Nx, *ytmp=yp+meff*Ny;
    cblas_dgemv(CblasRowMajor, CblasNoTrans, Ny, K, 1.0, Xtmp, K, beta, 1, 0.0, Xbeta, 1);    
    double fBuff=0.0;
    #pragma simd reduction(+:fBuff)
    for (int n=0; n<Ny; n++) {
      fBuff += loglike_base_f(Xbeta[n], ytmp[n]);
    }
    f += fBuff;
    delete [] Xbeta;
  }
  return f;
}
--------------------------------------------------------------------------
\end{verbatim}
\caption[]{NUMA-aware code for memory allocation and log-likelihood evaluation for logistic regression. For our test system, \texttt{Ncore} is 16. It is assumed that a compact processor affinity has been enforced so that threads 0 and 8 are each attached to a different processor. Alternatively, explicit affinity masks can be used for memory allocation on two sockets.}
\label{fig:code-numa-malloc}
\end{figure}

\begin{center}
*********
\end{center} 

\begin{figure}
\begin{centering}
\includegraphics[scale=0.65]{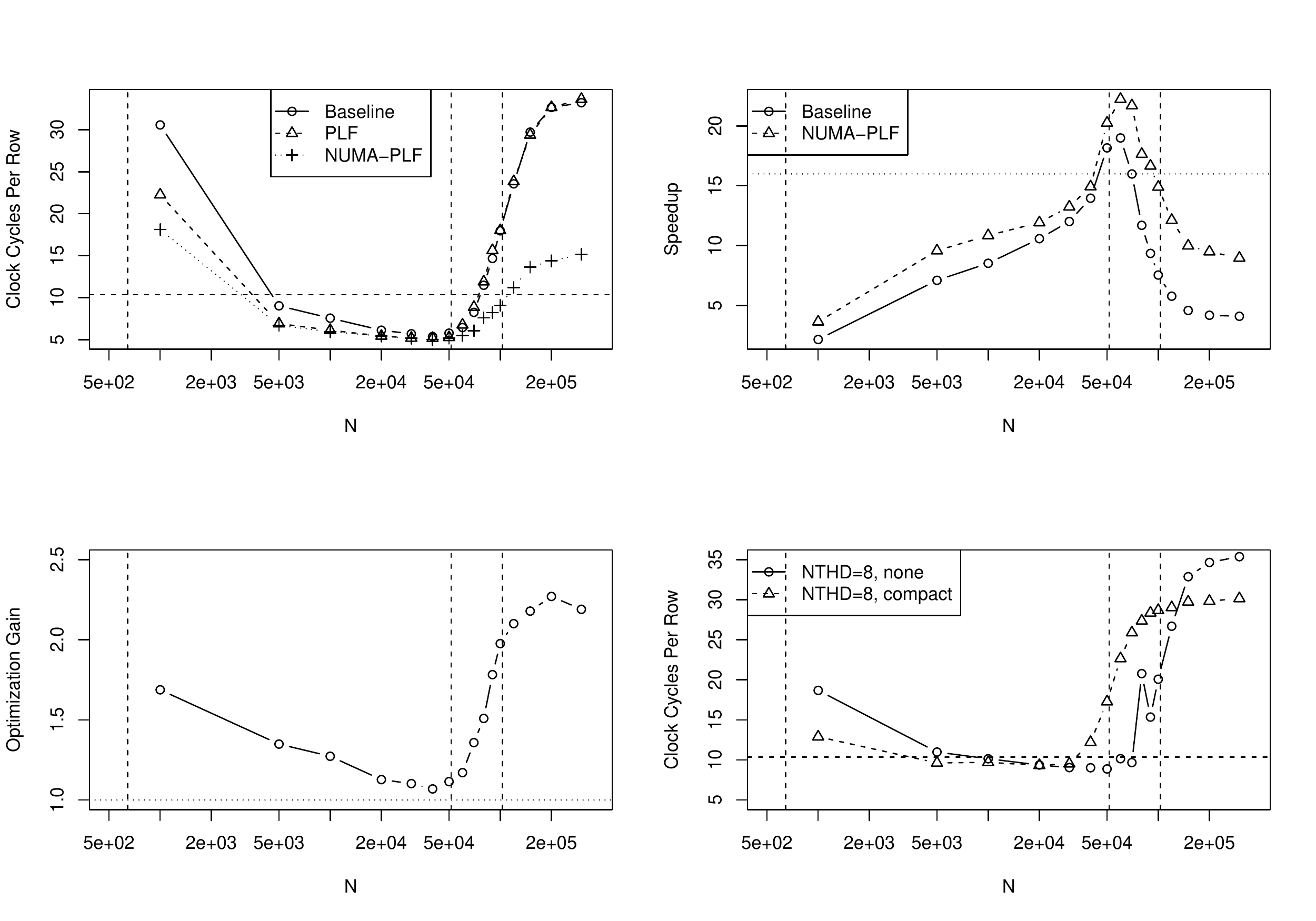}
\caption[]{Performance Optimization Results. Top Left: CPR for baseline, PLF, and NUMA-PLF implementations. Top Right: Speedup from 16 threads for baseline and NUMA-PLF code. Bottom Left: NUMA-PLF to baseline optimization gain as a function of $N$. Bottom Right: CPR for 8 threads, under two affinity policies. Vertical lines indicate L2 and L3 (per socket and total) cache boundaries. Horizontal lines indicate memory-based minimum CPR (top left and bottom right), linear speedup (top right) and no speedup (bottom left).}
\label{fig:opt-results}
\end{centering}
\end{figure}

Fig. \ref{fig:opt-results} shows the results of applying the above two performance tuning techniques, PLF and NUMA adjustments, in succession (labelled as `PLF' and `NUMA-PLF'). PLF reduces multi-threading overhead, leading to nearly 2x speedup for small data sizes, while NUMA adjustments also produce a nearly 2x speedup in the big-data regime. Our big-data performance is now within 46\% of memory-based hardware limit (top left, horizontal line).

Minimizing cross-socket memory transactions reinforces the assumption that task durations in the parallel region are nearly equal. This allows for a static thread scheduling to remain effective, and we can avoid the extra overhead associated with dynamic thread scheduling (Fig. \ref{fig:perfanalysis-1}, top left).

\subsubsection{Cache contention vs. Data Locality} \label{subsubsec-perfopt-cachecontention}
When using all 16 cores, we see clear benefit from increasing data locality by using a compact affinity policy in conjunction with NUMA-aware memory allocation. When using fewer than 16 cores, the picture is more complex, as seen in Fig. \ref{fig:opt-results} (bottom right). If total data size is larger than per-socket L3 but smaller than total L3, then spreading out the threads across sockets is advantageous since it allows for better utilization of total L3 (L3 contention regime). In contrast, when all data fits within L3 on each socket, keeping all cores on the same socket reduces inter-thread communication (L3 locality regime). When data is much bigger than total L3, it must be fetched from memory and the NUMA topology favors a compact affinity (NUMA locality regime). This tension between data locality and cache contention in NUMA multi-core systems has been noted in the literature (\cite{majo2011memory}).

\subsubsection{Differential Update} \label{subsubsec-perfopt-diffupdate}
The optimization strategies discussed so far have broad applicability, as they were focused on restructuring a given computation to minimize parallelization overhead and maximize data locality. Here we present another optimization technique, `differential update', that specifically applies to MCMC sampling. It seeks to remove unnecessary computation during Gibbs cycles and applies to graphs with continuous as well as discrete nodes. In the context of Bayesian GLM, it is more useful for univariate samplers such as slice sampler, and less useful for multivariate samplers such as Metropolis (\cite{metropolis2004equation}) or Hamiltonian Monte Carlo (HMC) (\cite{duane1987hybrid,neal1995bayesian}). It is also less useful for samplers that require evaluation of log-likelihood derivatives (Section \ref{subsec-deriv}). In this section we apply differential update to Bayesian GLM, and in Section \ref{subsec-beyond-glm} we discuss the concept for discrete graphs such as Boltzmann Machines and LDA.

Recall that in Gibbs sampling, we cycle through stochastic nodes and draw from the conditional distribution of each node, given the most recent samples for the remaining nodes. Using a univariate sampler such as the slice sampler (\cite{neal2003slice}), each of the $K$ elements of the coefficient vector $\bbeta$ are updated one at a time. So both inside the sampling algorithm as well as at the end of it, only one $\beta_k$ is being updated at a time, while the other components $\bbeta_{-k}$ remain fixed. The key insight in differential update technique is that we can update changes to $\XX \bbeta$ following a change to $\beta_k$ without having to perform the full computation. Instead, we recognize that
\begin{align}
\XX \, \bbeta &= [\XX_{-k} \,\, \XX_k] \left[ \begin{array}{c} \bbeta_{-k} \\ \beta_k \end{array} \right] \\
&= \XX_{-k} \, \bbeta_{-k} + \beta_k \XX_k,
\end{align}
where $\XX_{-k}$ is the result of removing column $k$, i.e. $\XX_k$, from $\XX$. From the above expression, we can see that a change to $\beta_k$ leads to the following change to $\XX \, \bbeta$:
\begin{equation}
\Delta (\XX \, \bbeta) = \Delta \beta_k \,\, \XX_k
\end{equation}

Our strategy, therefore, is to maintain a copy of $\XX \, \bbeta$ throughout the Gibbs sampling process, updating it according to the above formula after proposed or materialized changes to $\beta_k$'s. Fig. \ref{fig:code-diff-update} shows the resulting function. This strategy both reduces the amount of computation needed (in LA), and more importantly, reduces the data that must be read into memory nearly by a factor of $K$. For large $K$, this optimization has very significant impact.

\begin{figure}
\begin{verbatim}
--------------------------------------------------------------------------
double loglike(double dbetak, double Xbeta[], double Xt[], double y[], int k, int N, int K) {
  double f=0.0;
  #pragma omp parallel for num_threads(NTHD) reduction(+:f)
  #pragma simd reduction(+:f)
  for (int n=0; n<N; n++) {
    double xbetaTmp = Xbeta[n]+dbetak*Xt[k*N+n];
    f -= (log(1.0+exp(-xbetaTmp))+(1.0-y[n])*xbetaTmp);
  }
  return f;
}
--------------------------------------------------------------------------
\end{verbatim}
\caption[]{Log-likelihood function for binary logistic regression, using a differential update strategy. A separate function materializes the sampled $\beta_k$ into $\XX \, \bbeta$. To improve data locality, we use a transposed version of $\XX$ to align the data column-wise, referred to as \texttt{Xt} in the code. To avoid clutter, NUMA adjustments are not shown here but used in testing. Other routines are responsible for initializing \texttt{Xbeta}, and for updating it after each sampling step is finished.}
\label{fig:code-diff-update}
\end{figure}

\begin{figure}
\begin{centering}
\includegraphics[scale=0.65]{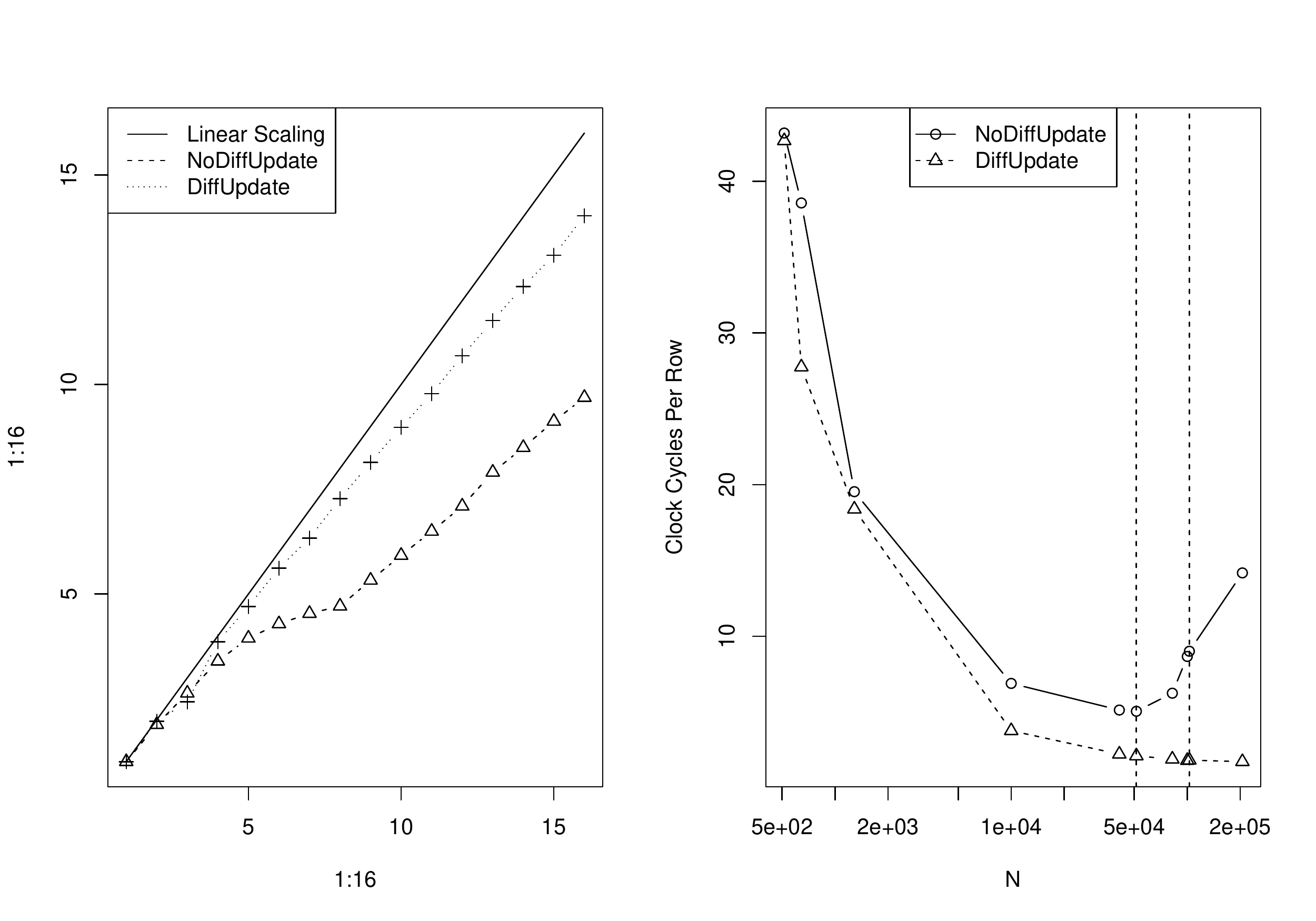}
\caption[]{Performance of differential update method for Bayesian logistic regression; $K=50$. Left: Speedup as a function of thread count, $N=200K$. Right: CPR as a function of $N$, 16 threads used.}
\label{fig:diff-update}
\end{centering}
\end{figure}

Fig. \ref{fig:diff-update} shows results for differential update. The left panel shows significantly better scaling of performance with number of cores for differential update method (`DiffUpdate'), compared to the default (`NoDiffUpdate') method for $K=50$ and $N=200K$. The right panel shows the impact of reduced data movement, where the differential update performance does not deteriorate when total data size exceeds L3 (in contrast to the NoDiffUpdate method). Essentially, DiffUpdate has expanded the range of $N$'s for which data fits within L3 by a factor of $K$. Fig. \ref{fig:perf-opt-diffmicro} compares micro-benchmarking results for DiffUpdate and NodiffUpdate for $K=50$, $N=200K$ (with vectorization and multi-threading on all 16 cores). We see that the overall speedup of $14.4/1.7=8.5$ can be broken down into a $1.2/0.5=2.4$x factor due to higher instruction execution rate (less memory stall), while the remaining $8.5/2.4=3.5$x is due to fewer instructions, i.e. reduced computation.

\begin{figure}
\begin{centering}
\begin{tabular}{|p{4cm}|p{4cm}|p{4cm}|}
\hline
 & \textbf{NoDiffUpdate} & \textbf{DiffUpdate} \\
\hline
Clock Cycles Per Row & 14.4 & 1.7 \\
\hline
L3 Hit Rate (\%) & 16.2 & 94.4 \\
\hline
Instructions Per Clock & 0.5 & 1.2 \\
\hline
\end{tabular}
\caption{Comparison of micro-benchmarking metrics between DiffUpdate and NoDiffUpdate methods for logistic regression. Parameters: $K=50$, $N=200K$, $\texttt{NTHD}=16$. For NoDiffUpdate, NUMA-PLF was applied. For DiffUpdate, NUMA adjustment was applied.}
\label{fig:perf-opt-diffmicro}
\end{centering}
\end{figure}

\begin{center}
**********
\end{center}

Fig. \ref{fig:perf-opt-summary} summarizes the impact of performance improvement techniques applied to the problem of calculating the log-likelihood for the logistic regression problem, using parameters $K=50$, $N=200k$. Note that in all cases we are using all 16 cores on the machine, and thus the performance improvement reflects more efficiency rather than simply using more resources.

\begin{figure}
\begin{centering}
\begin{tabular}{|p{4cm}|p{3cm}|p{7cm}|}
\hline
\textbf{Optimization Technique} & \textbf{CPR (Speedup)} & \textbf{Description} \\
\hline
Parallelized LA map & 116.9 (NA) & Multi-threaded MKL \\
\hline
Parallelized TR map & 34.3 (3.4x) & Use \texttt{omp} and \texttt{simd} pragmas \\
\hline
NUMA-PLF & 14.4 (8.1x) & Partial loop fusion and NUMA adjustments \\
\hline
Differential Update & 1.8 (64.9x) & Maintain and update \texttt{Xbeta} \\
\hline
\end{tabular}
\caption{Summary of performance optimization techniques applied to Bayesian logistic regression problem and their impact. Parameters: $K=50, N=200K$. 16 threads are used.}
\label{fig:perf-opt-summary}
\end{centering}
\end{figure}

While our focus was on logistic regression, same techniques are directly applicable to other type of GLM such as Poisson regression. While the detailed steps in the TR map change, the LA map stays the same, and so do the underlying concepts of minimizing parallel overhead and data movement.

\section{Extensions} \label{sec-ext}
In this section, we explore several extensions of the performance optimization concepts developed in the previous section in the context of Bayesian GLM.
\subsection{Hierarchical Bayesian GLM} \label{subsec-simd-embed}
The multi-level framework combines heterogeneous but similar units into a single model, allowing for information sharing without complete homogeneity (\cite{gelman2007data}). HB allows for consistent treatment of uncertainty and incorporation of prior knowledge in multi-level models, and has been applied to quantitative marketing (\cite{peter2005bayesian}), visual information processing (\cite{lee2003hierarchical}), medicine (\cite{babapulle2004hierarchical}) and changepoint analysis (\cite{carlin1992hierarchical}), among others.

HB GLM can be constructed from a GLM DAG such as Fig. \ref{fig:logit-dag} by replicating it $M$ times, with $M$ being the number of regression groups. Each unit has a distinct set of explanatory and target variables as well as coefficients. These $M$ DAGs are tied together by sharing the parent nodes of the group coefficients. Since $\bbeta_m$'s are not parents to each other, nor do they have common children (due to groups having distinct observations), then according to Theorem \ref{theorem:ci-parallelism} they are conditionally independent and can be sampled concurrently within a Gibbs sampling framework. Within each group, as with ordinary GLM, we can calculate the child-node contributions to the log-likelihood of each $\bbeta_m$ in parallel. The factorizable log-likelihood functions for each coefficient vector is given by:
\begin{equation}
L_m(\bbeta_m) = - \sum_{n=1}^{N_m} \left\{ (1-y_{m,n}) \xx_{m,n}^t \bbeta_m + \log(1+\exp(-\xx_{m,n}^t \bbeta_m)) \right\} \,\, \mathrm{with} \,\, m=1,\hdots,M.
\end{equation}
Full discussion of HB parallelization is beyond the scope of this paper (and subject of future research). Here we focus on a key question: Given the two parallel modes available (parallel sampling of group coefficients, and parallel calculation of log-likelihood function for each group), how should available hardware resources, i.e. vectorization and multi-core multi-threading, be allocated to each mode?

\textbf{Vectorization} In Section \ref{subsec-baseline} we discussed how to fully vectorize the log-likelihood function by adding the \texttt{simd} pragma before the TR loop in \ref{fig:code-baseline}, which produced a nearly 3x speedup (exact numbers depend on parameter values). Since the coarser parallelism mode does not lend itself to a SIMD form (due to presence of function calls and complex conditional code e.g. in the slice sampling routine), it cannot benefit from vectorization. We therefore continue to dedicate vectorization to parallel log-likelihood calculation, and see a near-multiplicative speedup as a result.

\begin{figure}
\begin{centering}
\includegraphics[scale=0.65]{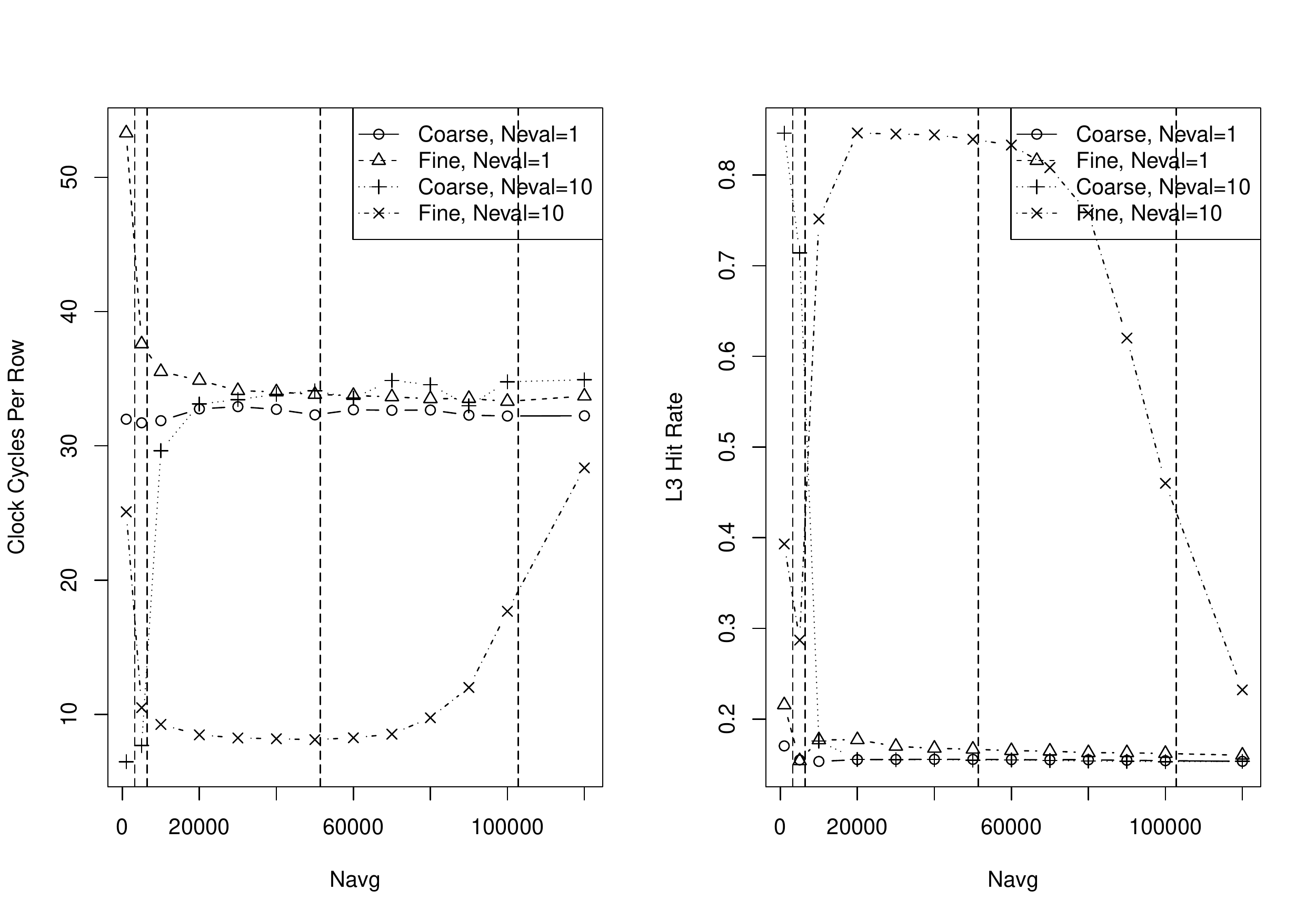}
\caption[]{Clock Cycles Per Row (left) and L3 hit rate (right) as a function of average group size, $Navg$, for HB logistic regression. `Coarse' means 16-core multi-threading is used for parallel sampling of regression groups, while 'fine' means 16-core multi-threading is used for parallel evaluation of log-likelihood within each group. \texttt{Neval} indicates the number of time each log-likelihood is evaluated. Two vertical lines to the right indicate values of $N$ beyond which average data size per regression group exceeds L3 cache size (per socket and total). Two vertical lines to the left indicate values of $N$ beyond which average data size multiplied by number of threads (16 here) exceeds L3 cache size (per socket and total). The latter numbers correspond to the average amount of data processed at any time by all threads under the coarse parallelism mode.}
\label{fig:hb-results}
\end{centering}
\end{figure}

\textbf{Multi-threading} How should the available cores be allocated to the two parallel modes? For example, should all 16 cores be dedicated to parallel sampling of coefficients, with log-likelihood calculation being single-threaded (but vectorized), or should we sample one group at a time but use a multi-threaded version of log-likelihood evaluator, or somewhere in-between such as 2x8, 4x4, or 8x2 arrangements? From the perspective of coverage and granularity (\cite{chandra2001parallel}), mapping coarse-grained parallelism to multi-threading is a better option since the longer duration of parallel tasks leads to better amortization of parallelization overhead. However, data locality considerations suggest a more complicated story, as seen in Fig. \ref{fig:hb-results}. We have compared the performance of two extreme resource allocation strategies for HB logistic regression problem with 100 regression groups and 50 covariates in each regression group, as a function of (uniform) group size ($Navg$). We have looked at two scenarios, one where the log-likelihood function for each group is evaluated only once per iteration ($\texttt{Neval}=1$), and one where it is evaluated 10 times ($\texttt{Neval}=10$). The latter is more representative of a real-world HB problem where each log-likelihood must be evaluated as often as a few hundred times for a univariate slice sampler. The interplay between \texttt{Neval} and mapping strategy is illuminating. When \texttt{Neval} is 1, data for each regression group is not reused until the next iteration, while when \texttt{Neval} is 10, data is reused and we see that mapping fine-grained parallelism to multi-threading is advantageous - by more than 3x - over a range of data sizes where each group fits in L3 but data size per group times \texttt{NTHD} exceed L3 capacity. The latter is the total amount of data processed by all threads at any given time, under the coarse-parallelism approach. This is supported by the measured L3 cache hit rates (right panel), where we see a much better cache reuse over the middle range of data when using multi-threading for SIMD parallel mode.

The broader lesson here is that, for data sizes and access patterns where cache-friendliness matters, locality considerations may favor using multi-threading for a fine-grained, cache-friendly parallelism over a coarse-grained but less cache-friendly parallelism. Similar situations to HB can arise in other settings, especially in the data analytics lifecycle. Examples are 1) variable selection using forward or backward selection (\cite{sutter1993comparison}) where many candidate explanatory variables are added to (or removed from) a base set and coarse-grained parallelism corresponds to parallel execution of the estimation algorithm against all candidate sets, 2) independent regressions on groups as a pre-cursor to HB, 3) execution of multiple models on various subsets of data and using different feature sets in the context of ensemble learning (\cite{polikar2006ensemble}). In all these cases, the coarse (task) parallelism is not cache-friendly, and a cache-friendly, fine-grained parallelization of the underlying algorithm can outperform task parallelism. This conclusion reinforces the case for continued investment by the HPC community in developing efficient, parallelized implementations of popular statistical and machine learning algorithms.

This discussion can be further generalized to the case where coarse-grained parallelism corresponds to parallel execution of multiple programs in a multi-user environment. In this case, the operating system is responsible for allocating processor cycles against all running applications, and has the unenviable task of striking a balance between maintaining data locality by offering long time slices to each process, and keeping the schedule `fair' by frequently switching context between processes (\cite{bovet2005understanding}).

A more thorough analysis of HB parallelization requires attention to other topics such as static-vs-dynamic thread scheduling and NUMA-aware memory allocation. For example, a primary application of HB framework is where some regression groups are so small that they need to borrow information from other, larger groups. Therefore, assuming that all groups have the same size is unrealistic. But non-uniform group sizes means non-uniform task durations, raising the possibility that a dynamic scheduling may be better than static scheduling. This, however, may have negative impact on data locality if cores are assigned to different groups each time. It is possible that the best strategy is to use a proxy for execution time as a function of group size and pre-allocate groups to threads to have even distribution and maintain a static mapping to preserve data locality. It must be noted that such potential complexities and performance hits associated with coarse-grained parallelism strengthen the case for using multi-threading with fine-grained parallelism. Further exploration of such issues is the subject of future research.

\subsection{Calculating Derivatives} \label{subsec-deriv}
In the context of MCMC sampling of probabilistic DAGs, speeding up the conditional log-likelihood function is highly impactful because the majority of time in a method such as Slice Sampler (\cite{neal2003slice}) or Metropolis (\cite{metropolis1949monte}) is spent on function evaluation. There exist other sampling techniques such as Adaptive Rejection (\cite{gilks1992adaptive}), HMC (\cite{neal1995bayesian}) and its variants such as Riemann Manifold HMC (\cite{girolami2011riemann}) or No-U-Turn Sampler (\cite{hoffman2011no}), and Hessian-based Metropolis-Hastings (\cite{qi2002hessian}), which use higher derivatives such as the gradient vector or the Hessian matrix. Similarly, in optimization it is advantageous to use the function derivatives when available (\cite{nocedal2006numerical}). Here we discuss parallel calculation of the gradient vector for GLM. Extension of this work to higher derivatives, namely Hessian, is the subject of future research.

Starting with Eq. \ref{eq-loglike-binlogit}, we use indexed functions $f_n$ absorb the dependence on $y_n$ and differentiate once to get the gradient vector:
\begin{align}
\g_L(\bbeta) &\equiv \frac{\partial L}{\partial \bbeta} \\
&= \sum_{n=1}^N f_n'(\xx_n^t \bbeta) \, \xx_n \label{eq-grad-expanded} \\
&= \XX^t \, \g_f, \label{eq-grad-compact}
\end{align}
where $\g_f \equiv [f_1'(\xx_1^t \bbeta) \hdots f_N'(\xx_N^t \bbeta)]^t$. The expanded form of Eq. \ref{eq-grad-expanded} makes the sum over contributions from each observation explicit, while the compact form of Eq. \ref{eq-grad-compact} hides this.

Fig. \ref{fig:code-fgh-compact} shows the baseline implementation of the routine for calculating the log-likelihood function and its gradient, following the SOM pattern. The routine now consists of a sequence of three maps: 1) calculate $\XX \bbeta$ (first LA map), 2) apply $f_n$ and $f'_n$ to each element of $\XX \bbeta$ (TR map), 3) calculate $\XX^t \g_f$ (second LA map). Note that for the last step we have switched to a column-major version of $\XX$ (referred to as \texttt{Xt} in code) for better data locality and parallelization scaling in the \texttt{dgemv} routine. Fig. \ref{fig:fgh} shows the performance of this baseline implementation. We can apply the same PLF and NUMA optimizations discussed in \ref{subsec-perfopt} here as well, with similar results. However, even after these two adjustments, we are quite off from the hardware limit. For example, Fig. \ref{fig:fgh} shows that for $N=300K$, after PLF and NUMA changes we have a CCR of $\sim$28, which is significantly higher than the memory-based minimum value of 10.4. This is somewhat expected since gradient calculation adds to the amount of computation per row. But we can do better, as explained below.

\begin{figure}
\begin{verbatim}
--------------------------------------------------------------------------
double loglike_fg(double beta[], double X[], double Xt[], double y[], int N, int K, double g[]) {
  double *Xbeta = new double[N];
  double *gf = new double[N];

  cblas_dgemv(CblasRowMajor, CblasNoTrans, N, K, 1.0, X, K, beta, 1, 0.0, Xbeta, 1);

  double f=0.0;
  #pragma omp parallel for num_threads(NTHD) reduction(+:f)
  #pragma simd reduction(+:f)
  for (int n=0; n<N; n++) {
    double u=1.0+exp(-Xbeta[n]);
    f += -(log(u)+(1.0-y[n])*Xbeta[n]);
    gf[n] = y[n]-1.0/u;
  }
  
  cblas_dgemv(CblasRowMajor, CblasNoTrans, K, N, 1.0, Xt, N, gf, 1, 0.0, g, 1);

  delete [] Xbeta;
  delete [] gf;
  return f;
}
--------------------------------------------------------------------------
\end{verbatim}
\caption[]{Multi-threaded, vectorized C implementation of the log-likelihood function and its gradient vector for logistic regression, using compact representation in Eq. \ref{eq-grad-compact}, leading to a SOM parallel pattern. A multi-threaded, vectorized MKL library is used for BLAS calls.}
\label{fig:code-fgh-compact}
\end{figure}

\subsubsection{Cache Fusion} \label{subsubsec-deriv-cachefusion}
We mentioned the primary advantage of MOS over SOM parallel pattern as one of consolidating the parallel regions and hence reducing thread management overhead. There is a related benefit in terms of minimizing data movement. In SOM, the output of each map stage is likely to be written to memory if array sizes are large, thus forcing the next map stage to reach down to the memory to retrieve the input it needs (assuming that the output of first map is input to second map). By transforming to MOS, we allow all operations to be chained and hence preventing the data from travelling unnecessarily to and from memory, or even cache since they will remain in registers. Our PLF strategy allowed us to minimize thread management overhead, but it did not achieve the same data locality as what would have been achieved in MOS. Cache fusion (\cite{mccool2012structured}) allows us to maintain better cache locality. The idea is to break the map into smaller chunks and execute them sequentially, thus making inputs and outputs of each stage smaller, possibly fitting within L1 or L2. This strategy offers measurable but small benefit for function evaluation (data not shown). But with the addition of derivatives, enhancing cache locality has a more significant impact since there are more steps and more computation. Fig. \ref{fig:code-fgh-hybrid} shows the code for function and gradient calculation with all three improvements (PLF, NUMA, Cache Fusion). We refer to the entire setup as Hierarchical Loop Structure (HLS), which can be a useful parallel pattern for other problems as well.

\begin{figure}
\begin{centering}
\includegraphics[scale=0.65]{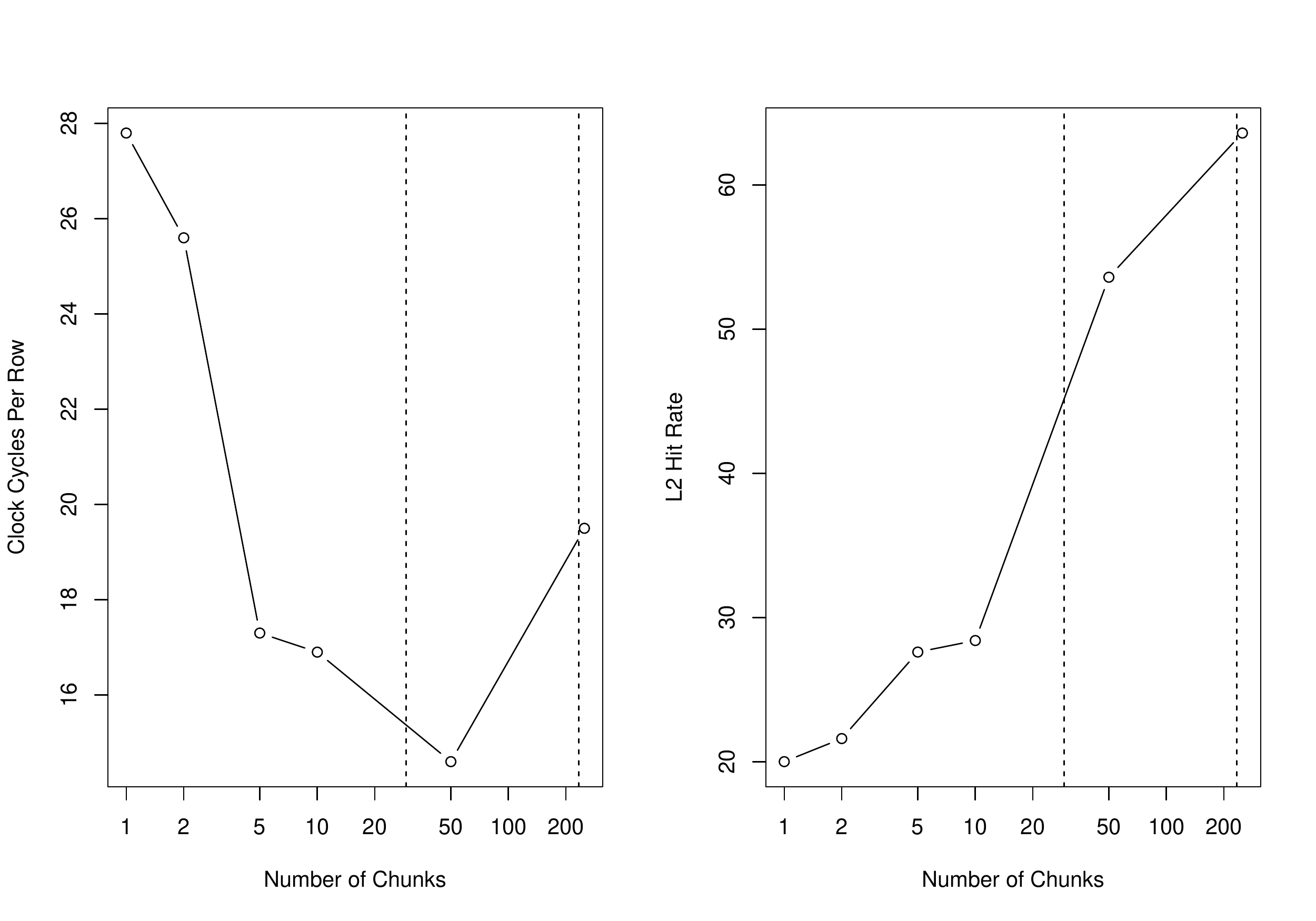}
\caption[]{Effect of `number of chunks' (\texttt{NCHUNK}) on CPR (left panel) and L2 cache hit rate (right panel), using the cache fusion strategy for logistic regression function and gradient calculation. Code is shown in Fig. \ref{fig:code-fgh-hybrid}. Vertical lines indicate values of \texttt{NCHUNK} beyond which each thread's share of data would fit within L2 (left vertical line) or L1 (right vertical line) cache.}
\label{fig:cache-fusion-micro}
\end{centering}
\end{figure}

Fig. \ref{fig:cache-fusion-micro} shows the impact of \texttt{NCHUNK} on performance of the code in Fig. \ref{fig:code-fgh-hybrid}, for $K=50$ and $N=300K$. For these problem dimensions, we see that: 1) we must choose the chunks small enough to fit within L2 to get the best performance (\texttt{NCHUNK=50}), but making the chunks too small has adverse impact on performance (left panel). The corresponding improvement in L2 hit rate can be seen in the right panel. Further research is needed to better understand the tradeoffs involved in reducing chunk size for L2 locality.

\begin{figure}
\begin{centering}
\includegraphics[scale=0.65]{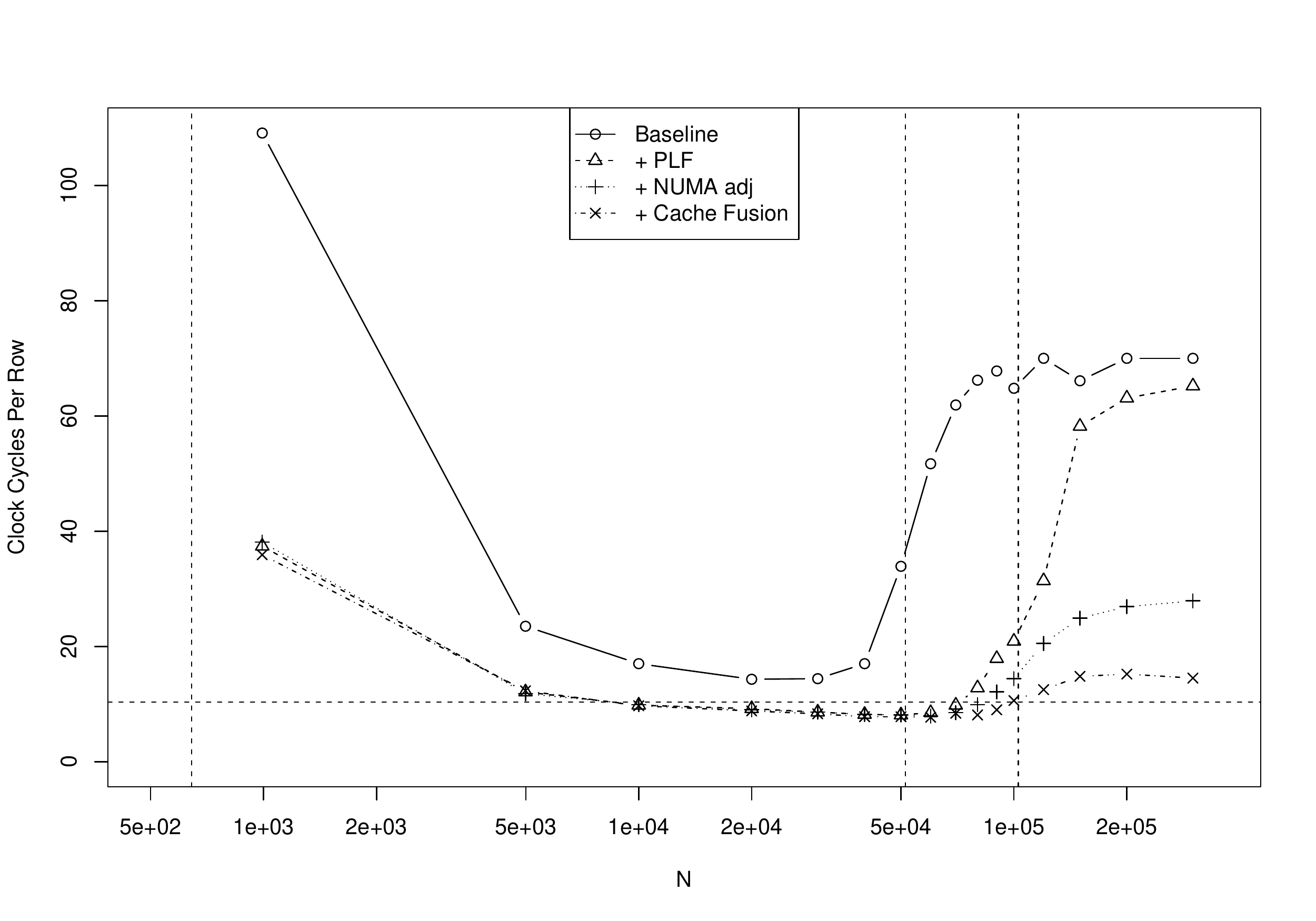}
\caption[]{Impact of successive code optimization techniques on performance of routine that calculates log-likelihood (return value) and its gradient vector (\texttt{g[]}) for logistic regression. $K=50$. Final code containing all optimizations in shown in Fig. \ref{fig:code-fgh-hybrid}. Chunk size selection for cache fusion is described in text. Vertical lines correspond to values of $N$ at which total data size exceeds L2 and L3 (per socket and total) cache size.}
\label{fig:fgh}
\end{centering}
\end{figure}

In Fig. \ref{fig:fgh} we have shown the impact of adding cache fusion to the previous optimization techniques, for $K=50$. For each value of $N$, we chose the largest chunk size that would fit within L2 (256KB) (also ensuring that is it a multiple of $\texttt{N/NTHD}$ for code simplification). We see that the impact is significant for large $N$, e.g. reducing CPR from 28 to 14.5, bringing it within 40\% of hardware limit (10.4).

\begin{figure}
\begin{verbatim}
--------------------------------------------------------------------------
double loglike_fg(double beta[], double X1[], double X2[], double y1[], double y2[]
  , int N, int K, double g[]) {
  memset(G, 0, K*sizeof(double));
  int Ny=N/NTHD, Nx=Ny*K, NTHD2=NTHD/2;
  double f=0.0;
  
  #pragma omp parallel num_threads(NTHD)
  { // start of parallel omp region
    int m=omp_get_thread_num(), meff;
    double *Xp, *yp;
    if (m<NTHD2) {
      Xp=X1; yp=y1;
      meff=m;
    } else {
      Xp=X2; yp=y2;
      meff=m-NTHD2;
    }
    double fBuff=0.0;
    double *ytmp=yp+meff*Ny, *Xtmp=Xp+meff*Nx;
    double *gtmp = new double[K];
    double *Xbeta = new double[Ny/NCHUNK];
    for (int i=0; i<NCHUNK; i++) { // start of cache fusion loop
      cblas_dgemv(CblasRowMajor, CblasNoTrans, Ny/NCHUNK, K, 1.0, Xtmp+i*K*(Ny/NCHUNK)
        , K, beta, 1, 0.0, Xbeta, 1);      
      #pragma simd reduction(+:fBuff)
      for (int n=0; n<Ny/NCHUNK; n++) {
        double r=1.0+exp(-Xbeta[n]);
        fBuff += -(log(r)+(1.0-ytmp[n+i*(Ny/NCHUNK)])*Xbeta[n]);
        Xbeta[n] = ytmp[n+i*(Ny/NCHUNK)]-1.0/r;
      }
      cblas_dgemv(CblasRowMajor, CblasTrans, Ny/NCHUNK, K, 1.0, Xtmp+i*K*(Ny/NCHUNK)
        , K, Xbeta, 1, 0.0, gtmp, 1);
      #pragma omp critical
      {
        for (int k=0; k<K; k++) {
          g[k]+=gtmp[k];
        }
      }
      f += fBuff;
    } // end of cache fusion loop

    delete [] Xbeta;
    delete [] gtmp;
  } // end of parallel omp region
  return f;
}
--------------------------------------------------------------------------
\end{verbatim}
\caption[]{C implementation of log-likelihood function and its gradient vector for logistic regression, with partial loop fusion, NUMA adjustments and cache fusion.}
\label{fig:code-fgh-hybrid}
\end{figure}

\subsubsection{Thread Synchronization} \label{subsubsec-deriv-sync}
In function evaluation, threads must synchronize while writing to the shared buffer holding the output. This was done through the \texttt{reduction} clause in the \texttt{omp} and \texttt{simd} pragmas in Figs. \ref{fig:code-baseline} and \ref{fig:code-expanded}. There is no equivalent construct for performing reduction on arrays (with the exception of FORTRAN OpenMP). We must instead use a \texttt{omp critical} pragma, as seen in Fig. \ref{fig:code-fgh-hybrid}. It is interesting to note that our PLF strategy for assigning jobs to threads has enormously reduced the overhead of this critical region, compared to the overhead of a brute-force MOS implementation shown in Fig. \ref{fig:code-fgh-expanded}. For example, we measured the penalty of the critical region for the PLF vs. MOS implementations for $N=10K$ and $K=50$, using 16 threads. For PLF, the critical region imposes an equivalent CPR overhead of 1.3, while for MOS, the same number is 475! Critical regions incur two types of penalty: first, the serialize the code and second, they require lock management by runtime.  Both these components are drastically reduced in the PLF approach since the critical region is only entered \texttt{NTHD} times, here being 16. In contrast, in MOS each job execution enters critical region once, amounting to $N$ times. As an alternative to a critical region, we can use \texttt{omp atomic} pragma before the addition statement inside the loop surrounded by the critical region construct in Fig. \ref{fig:code-fgh-expanded}. This slightly reduces the first penalty (due to shortening the serial code length) but increases the second penalty since mutex region is entered \texttt{K} times as many. Testing this on the MOS code, we see the penalty increasing from 475 to 1419. This suggests that lock management is the key performance factor in our synchronization operations. In contrast, a simple reduction of a scalar shows a much smaller performance hit when switching from PLF to MOS (0.2 CPR). Also, note that there is no support for vectorization of array reductions, and this is another blow to performance of the MOS implementation.

There is an equally important reason to avoid the MOS pattern of Fig. \ref{fig:code-fgh-expanded} as it relates to the last step (inside the critical region): The updates made to the array \texttt{g[]} by each thread will cause its cache lines to ping pong between cores \footnote{Note that this is not `false sharing' since threads do not simply share the same cache line, but are indeed writing to the same memory locations.}, creating another large performance hit. We validated this in two ways: 1) After removing the \texttt{omp critical} pragma (no synchronization), we measured CPR with and without the update line: \texttt{g[k]+=gtmp[k];}. We saw that this update line creates a 240-CPR penalty. 2) We measured L2/L3 hit rates and saw a drastic increase for both cache levels after removing the update line: 0.03\% $\rightarrow$ 71\% (L2) and 0.02\% $\rightarrow$ 95\% (L3). Again, the PLF approach works much better because each core does as much of the calculations privately, and updates the global \texttt{g[]} only once at the end. The cache hit rates for the code in Fig. \ref{fig:code-fgh-hybrid} agree with this story (76\% and 98\% for L2 and L3, respectively).

Finally, note that the differential update strategy will not be as effective when we need to calculate the gradient. This is because 1) calculating \texttt{Xbeta} is a smaller part of the entire computation, and 2) more importantly, differential update does not free us from the need to read the entire $\XX$ into cache during sampling for each $\beta_k$, since calculation of $\XX^t \g_f$ cannot be simplified to require only $\XX_k$.

\begin{figure}
\begin{verbatim}
--------------------------------------------------------------------------
inline double loglike_child_fg(double beta[], double x[], double y, int K
  , double g[]) {
  double xbeta = cblas_ddot(K, x, 1, beta, 1);
  double fbase, gbase, hbase;
  double r=1.0+exp(-xbeta);
  double fbase = -(log(r)+(1.0-y)*u), gbase = y-1.0/r;
  for (int k=0; k<K; k++) {
    g[k] = gbase*x[k];
  }
  return fbase;
}
double loglike_fg(double beta[], double X[], double y[], int N, int K
  , double g[]) {
  double f=0.0;
  #pragma omp parallel for num_threads(NTHD) reduction(+:f)
  #pragma simd reduction(+:f) /* despite the pragma, compiler will not vectorize this loop */
  for (int n=0; n<N; n++) {
    double gtmp[K];
    f += loglike_child_fg(beta, X+n*K, y, K, gtmp);
    #pragma omp critical
    {
      for (int k=0; k<K; k++) {
        g[k]+=gtmp[k];
      }
    }
  }
  return f;
}
--------------------------------------------------------------------------
\end{verbatim}
\caption[]{C implementation of log-likelihood function and its gradient vector for logistic regression, using expanded representation of Eq. \ref{eq-grad-expanded}, creating a MOS parallel pattern. A critical region is needed to sum contribution of each observation point towards gradient vector \texttt{g[]}. Presence of this critical region inside the for loop blocks its vectorization, despite the \texttt{simd} pragma.}
\label{fig:code-fgh-expanded}
\end{figure}

\begin{center}
**********
\end{center}

Fig. \ref{fig:opt-techniques-summary} summarizes the optimization techniques applied to parallel evaluation of log-likelihood and its gradient for Bayesian GLM, and the impact of each technique on the small-data and big-data problems. For big-data regime ($K=1250, N=500,000$), we reach within 25\% of the memory-bandwidth-induced CPR. This is 5x faster than our baseline implementation of Fig. \ref{fig:code-fgh-compact} (with or without TR parallelization), and 2.5x faster than the performance of Intel MKL's first \texttt{dgemv} call in the same code. NUMA adjustments explain this performance gap. Cache fusion provides the remaining 2x speedup, and impacts the chain of map sequences used in the routine by increasing L2 cache locality. Also note that small-data techniques (TR map parallelization and PLF) will be important when embedded in composite problems of larger size, such as HB GLM (Section \ref{subsec-simd-embed}).

\begin{figure}
\begin{centering}
\begin{tabular}{|p{4cm}|p{2cm}|p{2cm}|p{7.5cm}|}
\hline
\textbf{Optimization Technique} & \textbf{Small Data Speedup} & \textbf{Big Data Speedup} & \textbf{Explanation of Impact} \\ \hline
TR Map Parallelization & 2.4 & 1.01 & For wide data, LA map dominates computation, reducing the impact of TR parallelization. \\ \hline
Partial Loop Fusion & 2.5 & 1.02 & Total parallel overhead is independent of $N$, thus its amortized CPR is negligible for big data. \\ \hline
NUMA Adjustments & 1.1 & 2.5 & Small data fits in L3 of each socket, making memory bandwidth irrelevant. \\ \hline
Cache Fusion & 1.0 & 2.0 & For small data, each core's share fits in its L2, making cache fusion irrelevant. \\ \hline
\textbf{Overall Speedup} & \textbf{6.7} & \textbf{5.2} & Don't overlook small-data techniques for big-data problems: they become relevant in big-data composite problems such as HB. \\ \hline
\end{tabular}
\caption{Summary of optimization techniques for parallel evaluation of log-likelihood function and its gradient for logistic GLM. Small data: $N=10K$ and $K=10$. Big data: $N=500K$, $K=1250$. In all tests, 16 threads are used. Reference point is the baseline implementation of Fig. \ref{fig:code-fgh-compact}, WITHOUT the TR loop parallelized, i.e. the only source of parallelization are the \texttt{dgemv} BLAS calls.}
\label{fig:opt-techniques-summary}
\end{centering}
\end{figure}

\subsection{Bayond GLM} \label{subsec-beyond-glm}
So far, our discussion of SIMD parallel MCMC focused on Bayesian GLM. These models have two important characteristics: 1) lack of conjugacy between prior and likelihood for $\bbeta$ means we have no direct sampling technique available for the conditional posterior and must therefore use MCMC techniques whose core computational work involves repeated function and derivative evaluations, 2) the conditional posterior function is computationally expensive, scaling linearly with number of observations ($N$) as well as number of attributes ($K$) in the problem. Two implications follow from the above properties: 1) Random number generation does not constitute a significant fraction of sampling time, 2) the sampling routine is too complex for the coarse-grained parallelism (concurrent sampling of conditionally-independent nodes) to be vectorizable.

In this section, we look a different class of graphical models, Markov Random Fields (or undirected graphs) with discrete nodes. We will see that, despite several important differences between MRFs and GLMs, the PLF framework continues to be helpful, with a few new twists to consider.

\subsubsection{Ising Model} \label{subsubsec-beyondglm-ising}
We consider a square-lattice Ising model with binary nodes, with each node interacting with its 4 adjacent nodes. While Ising model has its origins in statistical mechanics (\cite{brush1967history}), it has been applied to statistical inference problems such as image reconstruction and denoising (\cite{perez1998markov,descombes1998spatio}) and neural information processing (\cite{hopfield1982neural}). The energy function for Ising model is given by:
\begin{equation}
E(\s) = \sum_i b_i s_i - \sum_{<i,j>} w_{ij} s_i s_j,
\end{equation}
where $\s$ is the state vector, $w_{ij}$ is the interaction weight between neighbors $i$ and $j$, and $b_i$'s control the bias terms. The sum occurs over pairs of $i$ and $j$ that are neighbors. In image de-noising, bias terms reflect the pixel intensity for the noisy image, serving to induce the output to represent the input. $w_{ij}$'s enhance correlation among neighboring pixels to reduce noise. The configuration probability is given by the Boltzmann distribution:
\begin{equation}
P(\s) = \frac{e^{-E(\s)}}{Z},
\end{equation}
where the normalizing factor $Z$ is called the partition function, and is given by $Z=\sum_{s}e^{-E(\s)}$. In Gibbs sampling of Ising model, the probability of node $i$ is given by:
\begin{equation}
P(s_i=1) = \frac{1}{1+\exp(-z_i)}, \label{eq-ising-p}
\end{equation}
where $z_i=b_i+\sum_{j} w_{ij}s_j$ and sum is over neighbors $j$ of node $i$.

\begin{figure}
\begin{centering}
\includegraphics[scale=0.65]{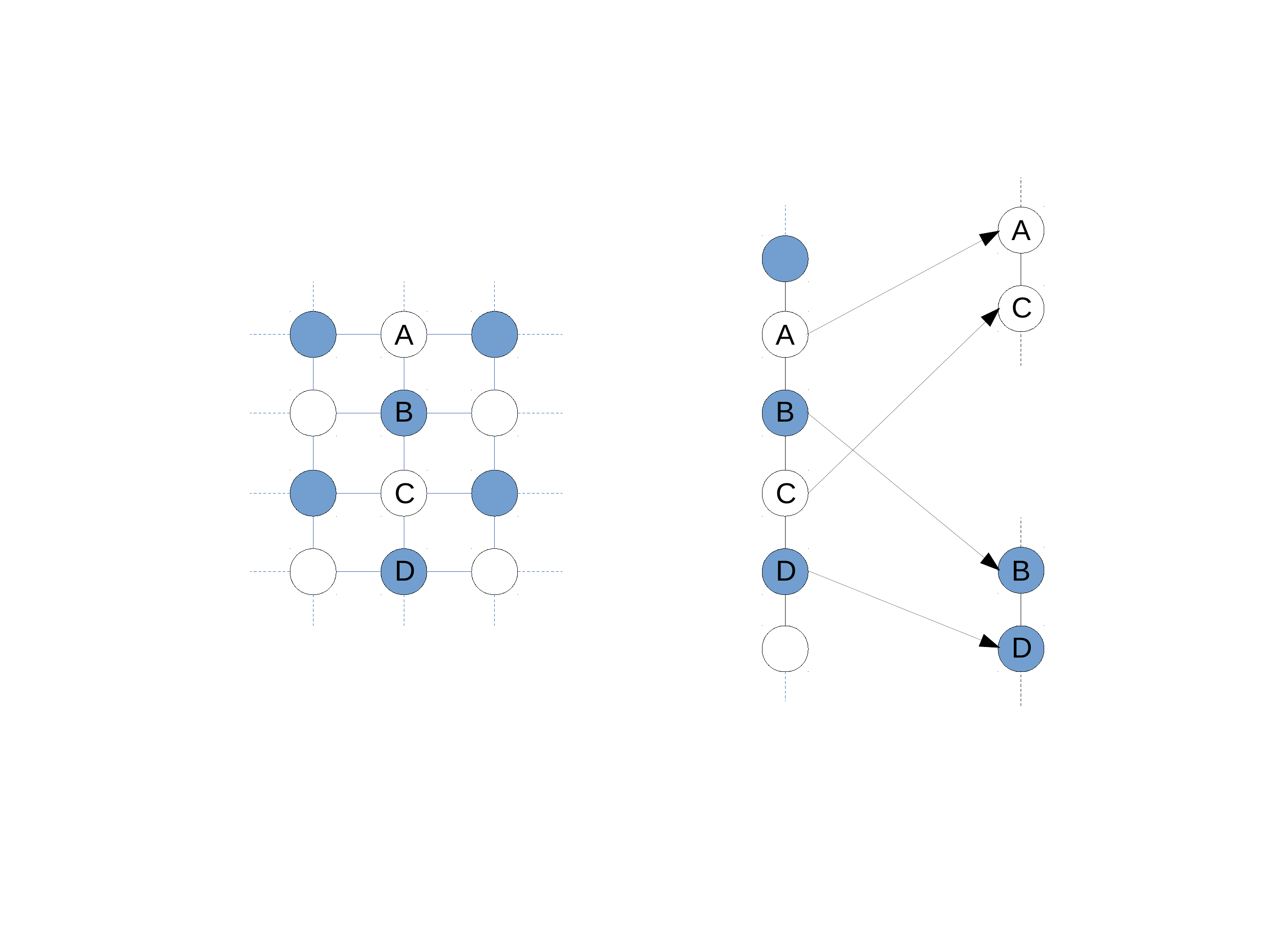}
\caption[]{Left: Illustration of graph coloring for lattice Ising model. All nodes of the same color are independent of each other, conditioned on the value of all nodes with the other color. Therefore, Gibbs sampling of this graph can proceed in two steps, in each of which all nodes with the same color can be sampled concurrently. Right: Default memory layout for the graph, showing that nodes of the same color will not occupy contiguous blocks of memory. However, an explicit mapping of each subset can create unit-stride memory access for Gibbs sampling of each colored subset.}
\label{fig:ising-graph}
\end{centering}
\end{figure}

This graph has a 2-color decomposition (Fig. \ref{fig:ising-graph}), and nodes within each color can be sampled concurrently. This is the coarse parallelism discussed in Section \ref{subsec-simd-gibbs}. On the other hand, since the sum in $z_i$ contains only 4 terms, the fine-grained parallelism opportunity is unattractive. It is clear, therefore, that multi-core multi-threading must be dedicated to the only available parallel mode. But the outstanding question is, can vectorization be applied here as well? Recall that in HB GLM, due to complexity of sampling steps for each $\beta_k$ and availability of fine-grained parallelism, we did not consider vectorization for coarse-grained parallelism. But in Ising model, the sampling steps for each node are significantly simpler, as described below:
\begin{enumerate}
\item Calculate $z_i$, the contribution from the neighbors. This consists of 4 multiplications and 5 additions. \label{step-ising-gather}
\item Calculate $P(s_i=1)$ according to Eq. \ref{eq-ising-p}. This involves 1 exponentiation, 1 division, and 1 addition. \label{step-ising-p}
\item Generating a sample for $s_i$. This consists of generating a random deviate from uniform distribution, and comparing it to the probability calculated in step \ref{step-ising-p}. \label{step-ising-rng}
\end{enumerate}
As always, the ideal scenario is to vectorize all steps. However, we face two challenges in doing so:
\begin{enumerate}
\item \textit{Memory Access Patterns:} In step \ref{step-ising-gather}, fetching neighbor states of a given node requires a gather memory operation. Handling boundary conditions adds further complexity to the code. Both of these can render vectorization inefficient or impossible.
\item \textit{RNG:} Making an external function call to a pre-compiled RNG library in step \ref{step-ising-rng} can act as a vectorization blocker.
\end{enumerate}
Here again we can take advantage of the PLF strategy for incremental vectorization. Given the memory access pattern challenge of step \ref{step-ising-gather} and its small contribution to total computation per sample, we choose to leave it scalar. As for step \ref{step-ising-rng}, we propose a batch RNG approach.

\textit{Batch RNG} Batch generation of uniform random deviates used in step \ref{step-ising-rng} offers two advantages. Firstly, it removes RNG call from the sampling loop, allowing it to be vectorized more efficiently. Secondly, it speeds up RNG by reducing the function call overhead and allowing for use of high-performance libraries that utilize vector instructions or other forms of parallelization. Examples include Intel VSL \footnote{\url{http://tinyurl.com/pevz69r}}, Nvidia's cuRAND \footnote{\url{https://developer.nvidia.com/curand}}, and the SPRNG project \footnote{\url{http://www.sprng.org/}}. Applying the batch RNG strategy is more involved for complex distribution. See Section \ref{subsec-batch-rng}.

\textit{Data Layout Optimization} Vectorization efficiency for step 2 depends on data alignment for $s_i$ (and $b_i$). If the entire network occupies a contiguous block of memory, then data for each (colored) graph partition will not be contiguous. Such non-unit-stride memory accesses have a particularly negative impact on vectorization. To overcome this, we can extract each sub-graph and write it to new arrays, perform Gibbs sampling on these contiguous-memory arrays, and transform the results back to the original arrangement at the end (Fig. \ref{fig:ising-graph}).

\begin{figure}
\begin{centering}
\includegraphics[scale=0.65]{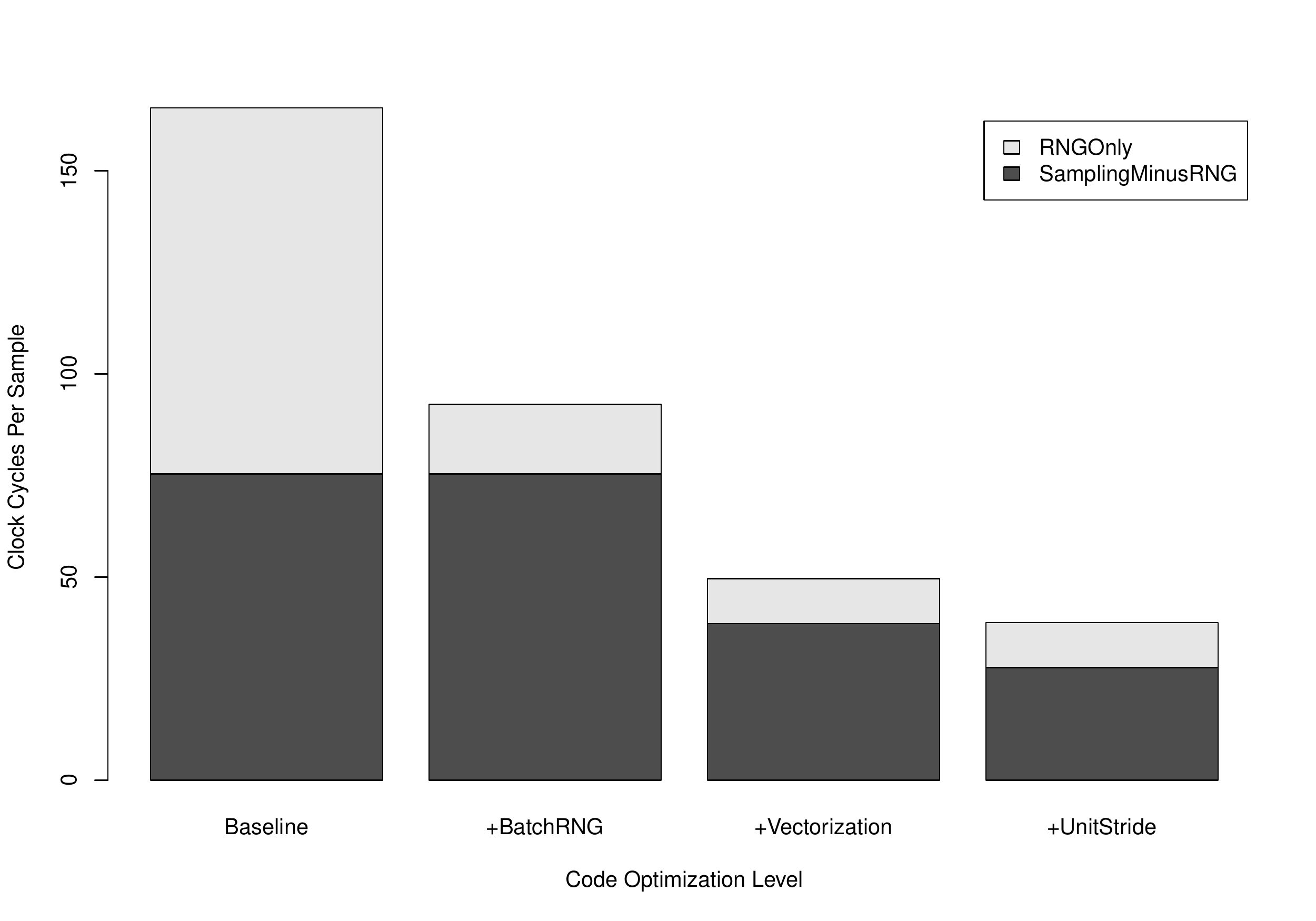}
\caption[]{Cumulative impact of three optimization techniques on performance of Gibbs sampling for a square-lattice Ising model used in image de-noising. Image dimensions are 560 by 516. 1000 samples were generated. The energy function followed the specification and parameters in Chapter 8 of \cite{christopher2006pattern}. First, we switched from one-at-a-time to batch RNG. Second, sampling step \ref{step-ising-p} is vectorized. Third, graph nodes belonging to each color are copied to contiguous blocks of memory, allowing for unit-stride access during vectorization.}
\label{fig:ising-progression}
\end{centering}
\end{figure}

Fig. \ref{fig:ising-progression} shows the impact of 3 optimization techniques on performance of Gibbs sampling for an Ising model used in denoising a 560-by-516 binary image (1000 samples). We followed the specification and parameters in \cite{christopher2006pattern} (Chapter 8). Switching to batch RNG provides a 4.9x speedup for the RNG part of the code (84.1 $\rightarrow$ 17.1 clock cycles per sample). Vectorization results in a 2.3x speedup in the non-RNG part of the code (75.4 $\rightarrow$ 32.5). Finally, switching to unit-stride memory access led to a 1.5x speedup in non-RNG part (32.5 $\rightarrow$ 21.7). The cumulative impact of all 3 optimization techniques (on the entire code) is a 4.3x speedup.

\subsubsection{Boltzmann Machine} \label{subsubsec-beyondglm-boltzmann}
A close cousin of Ising model is the Boltzmann machine (\cite{ackley1985learning}), where nodes are densely connected rather than having the lattice connection of Ising model. As a result, 1) there is no conditional independence among nodes and hence the graph cannot be colored for parallel sampling, 2) each node is potentially connected to a large number of nodes, creating a SIMD parallelization opportunity for calculation of $z_i$'s. A variant of Boltzmann machine is the Restricted Boltzmann machine (RBM) (\cite{smolensky1986information}), a key ingredient in building deep belief nets (\cite{hinton2006fast}) that are seeing increasing application in supervised and unsupervised learning of generative models for image (\cite{krizhevsky2012imagenet}), video (\cite{le2011learning}) and speech (\cite{mohamed2011deep}) data. RBMs have a bipartite graph structure, where visible and hidden nodes are densely connected, but no connection amongst nodes of each layer is allowed. This leads to conditional independence of hidden nodes, allowing for their parallel sampling. In choosing between the fine- and coarse-grained parallel modes to apply vectorization, we must consider these factors: 1) Fine-grained parallel mode consists of a summation over contribution from nodes connected to a given node. Its arithmetic complexity is low, but vectorization is easy. 2) Coarse-grained parallel mode contains the inner, summation loop. Efficient vectorization of this step within the coarse-grained parallelism would require loop unroll (see Section \ref{subsubsec-jit-unroll}). 3) Steps \ref{step-ising-p} and \ref{step-ising-rng} from the Ising mdoel are repeated here, and we can use strategies discussed before. Note that dense connectivity can alleviate or remove the problem of suboptimal memory access patterns.

\subsubsection{Differential Update and Atomic Vector Operations} \label{subsubsec-atomic-vector}
We introduced the differential update strategy in the context of Bayesian GLM models (Section \ref{subsec-perfopt}). This strategy can also be used with discrete-node graphs such as Ising models, Boltzmann machines and LDA models. For example, consider the image de-noising example introduced in this section. We measured transition rates for binary pixels in the model, finding that after the first few iterations less than 10\% of the pixels switch value. If we switch to a differential update approach where we maintain intermediate data structures $\mathbf{z}$ and $\mathbf{p}$, and update them each time a node has a state transition, we see a nearly 2x speedup. However, combining this differential update approach with vectorization can be a challenge. If two nodes are updated simultaneously and they share neighbors, they will attempt to update the same elements of $\mathbf{z}$ and $\mathbf{p}$, requiring atomic operations inside vectorized code. Unfortunately, current versions of x86 processors do not support atomic operations (\cite{kumar2008atomic}), and it is likely that future support will be rather inefficient. Similar opportunities and challenges exist for RBMs where conditional independence allows hidden nodes to be updated concurrently, but they could attempt to write to same locations in intermediate arrays. In non-collapsed Gibbs sampling of LDA, a differential update strategy is commonly used where matrices representing counts of words by topic and topics by document are maintained and updated after each new topic assignment for a token (\cite{xu2011multicore}). Vectorization of parallel sampling for topic assignments will face the same challenge of needing to perform atomic operations on these count matrices.

\subsection{Batch RNG for Complex Distributions} \label{subsec-batch-rng}
In Section \ref{subsec-beyond-glm} we saw that using a batch RNG for uniform deviates in Gibbs sampling of Ising model led to a nearly 2x speedup (Fig. \ref{fig:ising-progression}) and paved the way for vectorization of concurrent sampling of nodes of the same color. Distributions such as uniform and normal are easy to handle in batch process due to their linear transformation properties, allowing a deviate with standard parameters to be converted to a deviate of arbitrary parameters:
\begin{equation}
X \sim U(0,1) \longrightarrow a+(b-a)X \sim U(a,b),
\end{equation}
where $U(a,b)$ is the uniform distribution on the real interval $[a,b)$. Similarly, for the normal distribution we have:
\begin{equation}
X \sim N(0,1) \longrightarrow \mu + \sigma X \sim N(\mu,\sigma),
\end{equation}
where $N(\mu,\sigma)$ is the normal distribution with mean $\mu$ and standard deviation $\sigma$. If a distribution does not enjoy such properties, we cannot pre-generate its random deviates in advance without knowing the value of parameters needed. The best we can do is to pre-generate the `component' random deviates that are used in the algorithm for sampling from the desired distribution. As a concrete example, consider the Dirichlet distribution that is used in LDA models as a conjugate prior to categorical and multinomial distributions of topics in documents and words in topics (\cite{blei2012probabilistic}):
\begin{equation}
f(x_1,\hdots,x_K;\alpha_1,\hdots,\alpha_K) = \frac{1}{\mathrm{B}(\mathbf{\alpha)}} \prod_{i=1}^K x_i^{\alpha_i-1},
\end{equation}
where $x_i \in [0,1), \,\, i=1,\hdots,K$, $\sum_i x_i = 1$, and $\alpha_i>0, \,\, i=1,\hdots,K$. Dirichlet deviates can easily be constructed from Gamma deviates \footnote{First, we draw $y_i$'s from $\mathrm{Gamma}(\alpha_i,1)$. Next we normalize: $x_i = y_i/\sum_i y_i$.}, which in turn require uniform (and possibly normal) deviates. We took the Gamma sampling algorithm described in \cite{press2007numerical} \footnote{The algorithm uses the rejection method of \cite{marsaglia2000simple} for $\alpha>1$, and a lemma described in \cite{ripley2009stochastic} to cast an $\alpha<1$ problem into an $\alpha>1$ problem.}, and adapted it to use a buffer of uniform and normal random deviates, generated in batch. This Gamma sampler was then used in the Dirichlet sampler. For reference, we used a baseline Gamma sampler which generated its uniform and normal deviates one at a time. In both versions, Intel VSL routines were used for uniform and normal deviates. We tested the batch and one-at-a-time versions of Dirichlet sampler on the Enron email data set from UCI repository (\cite{BacheLichman2013}), specifically for sampling $\theta$, the probability of topics (100) for each document (39861). A flat (Dirichlet) Jeffreys prior was used for all documents. One-at-a-time Dirichlet sampler had a performance of 1263 clock cycles per Gamma sample, while the batch version had a performance of 198 clock cycles, showing a speedup of 6.4x. Fig. \ref{fig:batch-vs-oaat} summarizes the speedup offered by batch vs. one-at-a-time RNG for uniform, normal, Gamma, and Dirichlet distributions.

\begin{figure}
\begin{centering}
\begin{tabular}{|p{2cm}|p{3cm}|p{3cm}|p{3cm}|}
\hline
\textbf{Distribution} & \textbf{One-At-A-Time (CCPS)} & \textbf{Batch (CCPS)} & \textbf{Batch (Speedup)} \\
\hline
Uniform & 76 & 18 & 4.2 \\
\hline
Normal & 949 & 39 & 24.3 \\
\hline
Gamma & 1170 & 162 & 7.2 \\
\hline
Dirichlet & 1263 & 198 & 6.4 \\
\hline
\end{tabular}
\caption{Comparison of one-at-a-time and batch RNG methods, for 4 probability distributions. For uniform, normal and Gamma distributions, 10 million random deviates were generated. For Gamma, shape and rate parameters used are 2.0 and 3.0, respectively. For Dirichlet, we generated samples in the context of a LDA model according to the text. CCPS: Clock Cycles Per Sample. For Dirichlet, in order to normalize for the effect of $K$, CCPS is calculated per each Gamma sample generated. For uniform and normal deviates, Intel VSL library was used with methods being \texttt{VSL\_RNG\_METHOD\_UNIFORM\_STD} and \texttt{VSL\_RNG\_METHOD\_GAUSSIAN\_BOXMULLER}, respectively. For Gamma deviates, we used the algorithm described in \cite{press2007numerical}. For Dirichlet, we use the standard method of using Gamma deviates. Batch numbers include time for copying data from RNG buffer to destination, which imposes around 6 clock cycles per sample in overhead. The numbers for batch Gamma and Dirichlet include roughly a 10\% additional overhead for generating extra uniform and normal deviates since the exact number of base deviates needed per Gamma sample is not known in advance. A more optimize code can re-fill buffer periodically to reduce waste.}
\label{fig:batch-vs-oaat}
\end{centering}
\end{figure}

Our results suggest that there is value in RNG libraries offering batch generation capabilities. In the above example, we could not use the Gamma sampler offered by Intel VSL since its interface does not offer an option of passing buffers of uniform or normal deviates, forcing us to write the Gamma sampler from scratch.

\subsection{Beyond Single-Node x86} \label{subsec-manycore}

The focus of this paper has been on maximizing performance of a single, shared-memory x86 compute node for parallel MCMC sampling. Going beyond single-node x86 can be driven by three needs: 1) more memory, 2) more memory bandwidth, and 3) more computing power. In distributed-memory clusters, total memory and compute power scale linearly with node count, and so does total memory bandwidth between each node's cores and its private memory. However, typically-slower inter-node connections mean inter-node costs of synchronization and data exchange are higher than their intra-code counterparts. As such, the two principles of minimizing parallel overhead, and minimizing data movement must be followed more rigorously in distributed computing. In fact, we can consider our NUMA adjustments, discussed in Section \ref{subsec-perfopt}, as a precursor to distributed computing. Asymmetric treatment of cross-socket vs. within-socket communication, while continuing to take advantage of the convenience of shared-memory programming, can be considered halfway between writing symmetric, shared-memory parallel code and the standard message-passing-based approach to distributed computing on a cluster.

Many-core architectures such as Intel Xeon Phi and GPUs offer another option for adding memory bandwidth and compute power. They can be used as an alternative to, or in conjunction with, distributed computing. They are best suited for floating-point SIMD parallelism since they dedicate more transistors to floating-point operations and less to out-of-order execution, branch prediction, and speculative execution (\cite{kirk2012programming,jeffers2013intel}). Intel Xeon Phi co-processors can be considered an extension of x86 processors with higher core count (up to 60) and wider vector units (512-bit, holding 8 doubles). Fig. \ref{fig:gpu-cpu-compare} compares the compute and memory performance of an Intel Xeon Phi 7120P and an Nvidia Tesla K40 against our x86 test system. Based on these numbers, we can say, in broad terms, that compute-bound problems can gain $>$3.5x speedup while memory-bound problems can gain between 2x-3.5x in speedup by switching from the an x86 processor to a many-core co-processor. As with x86, approaching these hardware limits requires tuning to minimize parallel overhead and maximize data locality.

\begin{figure}
\begin{centering}
\begin{tabular}{|p{4cm}|p{3cm}|p{3cm}|p{3cm}|}
\hline
 & Intel Xeon E5-2670 (dual-socket) & Intel Xeon Phi 7120P & Nvidia Telsa K40 \\ \hline
Memory Bandwidth (GB/sec) & 102.4 & 352 (3.44x) & 288 (2.81x) \\ \hline
Peak Floating-Point Performance (GFLOPS/sec) & 332.8 & 1208 (3.63x) & 1430 (4.30x) \\ \hline
Memory (GB) & (up to) 384 & 16 (1:24) & 12 (1:32) \\ \hline
\end{tabular}
\caption{Comparison of dual-socket Intel E5 2650, Intel Xeon Phi 7120P and Nvidia Tesla K20 in terms of memory bandwidth and peak floating point performance. Numbers in parentheses indicate speedup of co-processor over the x86 reference system (2 x Intel Xeon E5-2650) for the given dimension. Sources: \url{http://tinyurl.com/p47ytt3} (intel), \url{http://tinyurl.com/oexotqv} (Nvidia)}
\label{fig:gpu-cpu-compare}
\end{centering}
\end{figure}

There are two other factors beyond the raw performance of co-processors that must be considered in the cost/benefit analysis of using co-processors (also referred to as `devices'): 1) memory capacity, 2) data transfer and parallelization overhead. As we see in Fig. \ref{fig:gpu-cpu-compare}, memory capacity is significantly lower for current generation of co-processors, compared to modern x86 servers. Therefore, as data sizes exceed device memory, the device may become significantly more complex to accommodate the need to move data in and out of device memory. In our GLM example, at $K=50$, the Tesla K40 card will run out of space for $N$ larger than 31.6 million rows, while the same limit on the host system will be reached at data sizes 32 times larger.

\begin{figure}
	\begin{subfigure}[b]{0.3\textwidth}
	\begin{verbatim}
------------------------------------------
for (int i=0; i<1000; i++) {
  cudaMemcpy(betaDev, beta, K*sizeof(double)
    , cudaMemcpyHostToDevice);
  doNothing<<<1,1>>>();
  cudaMemcpy(&f, fpDev, 1*sizeof(double)
    , cudaMemcpyDeviceToHost);
}
------------------------------------------
	\end{verbatim}
	\end{subfigure}
\hfill
	\begin{subfigure}[b]{0.5\textwidth}
\includegraphics[width=\textwidth]{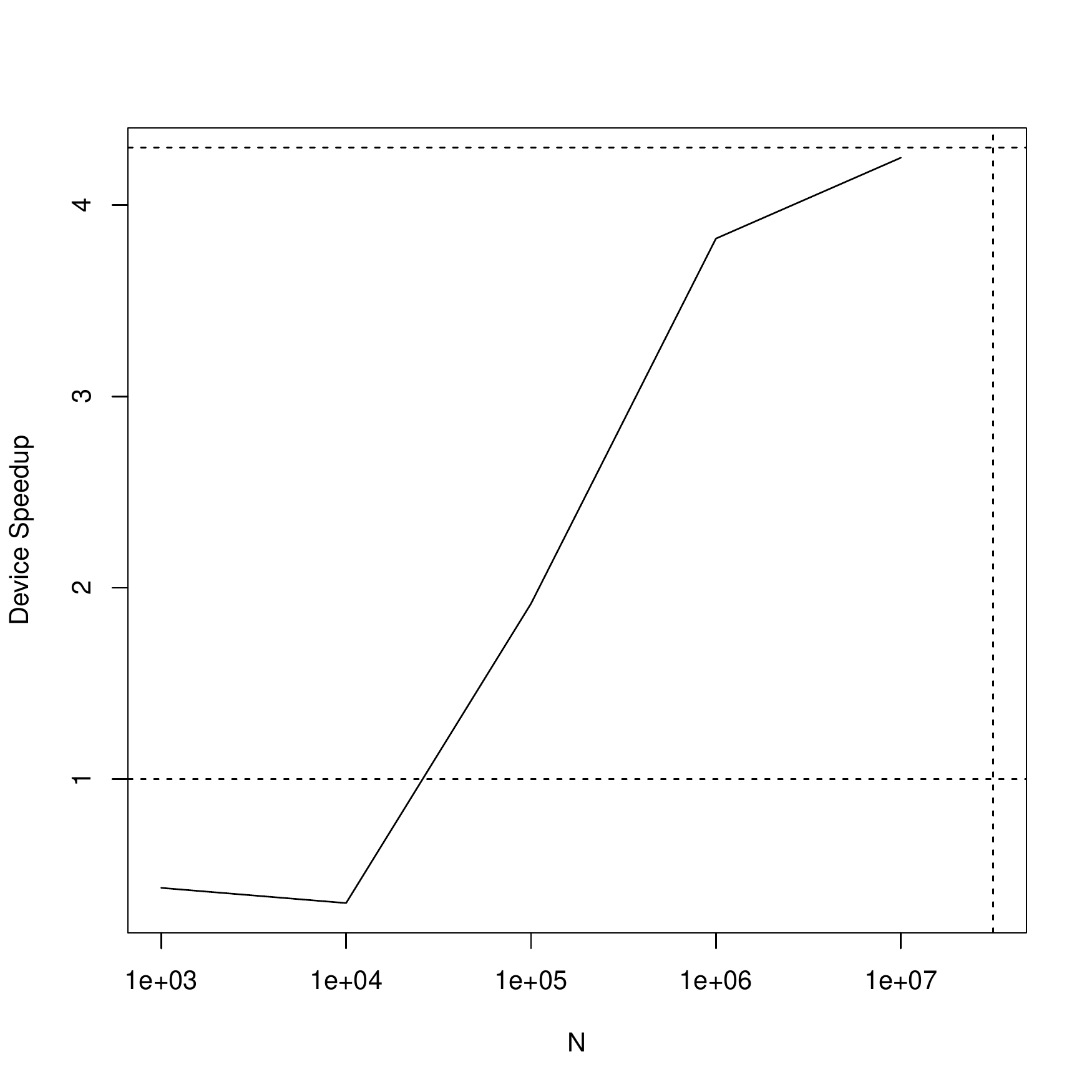}
	\end{subfigure}

	\caption{Left: CUDA code used to measure kernel launch and data transfer overhead for Bayesian GLM, assuming incomplete code port. Right: Expected speedup from porting log-likelihood function to Nvidia Tesla K20, as a function of $N$, for $K=50$. We assume both host and device code are compute-bound, hence a 4.3x maximum speedup (horizontal line) for device from Fig. \ref{fig:gpu-cpu-compare}, which is speedup assuming full code port. Using code in left panel, we measured overhead associated with incomplete code port to be a constant 20$\mu$sec, amortized over $N$ to produce an effective CPR overhead. Vertical line represents memory boundary for device, beyond which it cannot host $\XX$ in its 12GB global memory.} \label{fig:gpu-analysis}
\end{figure}

Unless the code is ported completely to the device, so that there is no need to send data back and forth between the host and device memory, we will incur data transfer cost, due to non-zero latency and finite bandwidth. On GPU cards, kernel launch incurs an additional cost, while on Xeon Phi where we may launch more than 100 threads, the thread management overhead can be significantly higher than the numbers we discussed for the host system. For example, in Bayesian GLM we may want to use the device only for function evaluation and run the sampler on the host, or there might be other parameters besides $\bbeta$ that are sampled. As a result, each function evaluation may incur the data transfer overhead associated with 1) sending $\bbeta$ from host to device, 2) launching the kernel on the device, and 3) sending the log-likelihood result from device back to the host. To measure all these overhead components, we used an empty kernel inside a for loop, as seen in Fig. \ref{fig:gpu-analysis} (left). (This test was performed on a Tesla K20 card.) This loop took nearly 20$\mu$sec to run per iteration, equal to $52K$ CPU clock cycles. If $N=10K$, this translates into a CPR overhead of 5.2. Recall that, using the differential update approach on the CPU we reached a CPR of less than 2.0 using all 16 cores (Fig. \ref{fig:perf-opt-summary}). Therefore, even ignoring the memory and compute limits of the device, simply the data transfer and kernel launch remove any incentive to port the code to the device, unless port a sufficient portion of the code to remove the need for such frequent kernel launches and data transfers. Fig. \ref{fig:gpu-analysis} (right) shows the expected speedup as a result of device hardware limits as well as kernel launch and data transfer overhead. This analysis can be used in making an informed decision on whether or not to port the code (and how much of it) to the device.

Finally, development and maintenance costs associated with code complexity must be factored into a total cost of ownership analysis of using co-processors. CUDA \footnote{\url{http://www.nvidia.com/object/cuda_home_new.html}} is a convenient programming language for  Nvidia GPUs, yet it is not open-source, while OpenCL \footnote{\url{https://www.khronos.org/opencl/}} has the opposite pros and cons. Even CUDA code for logistic regression log-likelihood will inevitably be more complex than the CPU code. For example, in CUDA kernels threads within a block can communicate inside a kernel, while blocks cannot. On the other hand, there is an upper limit on number of threads that can be launched within a block. Such limitations may force us to break up the CUDA kernel into two steps, one in which a partial reduction is performed within blocks, and a second kernel where the final sum is calculated. Optimal selection of grid and block geometry is another design decision that may require runtime and/or compile-time code branching.

\subsection{Compile-Time Loop Unroll} \label{subsubsec-jit-unroll}
In implementing the GLM log-likelihood, we encountered a nested loop in the LA map whose vectorization in the SOM and PLF patterns was delegated to Intel MKL library as part of the \texttt{dgemv} call (see Figs. \ref{fig:code-baseline} and \ref{fig:code-hybrid}). We also mentioned that, to transform this log-likelihood function to a MOS pattern, we can fuse the outer loop of the LA map with the TR loop, which leaves us with a \texttt{ddot} call inside the fused loop to represent the LA step (Fig. \ref{fig:code-expanded}). But the presence of this function call prevents a full vectorization of the outer loop, creating bad performance. This can be verified by inspecting the assembly code generated by the Intel compiler, as described below.

\begin{figure}
\begin{centering}
\includegraphics[scale=0.65]{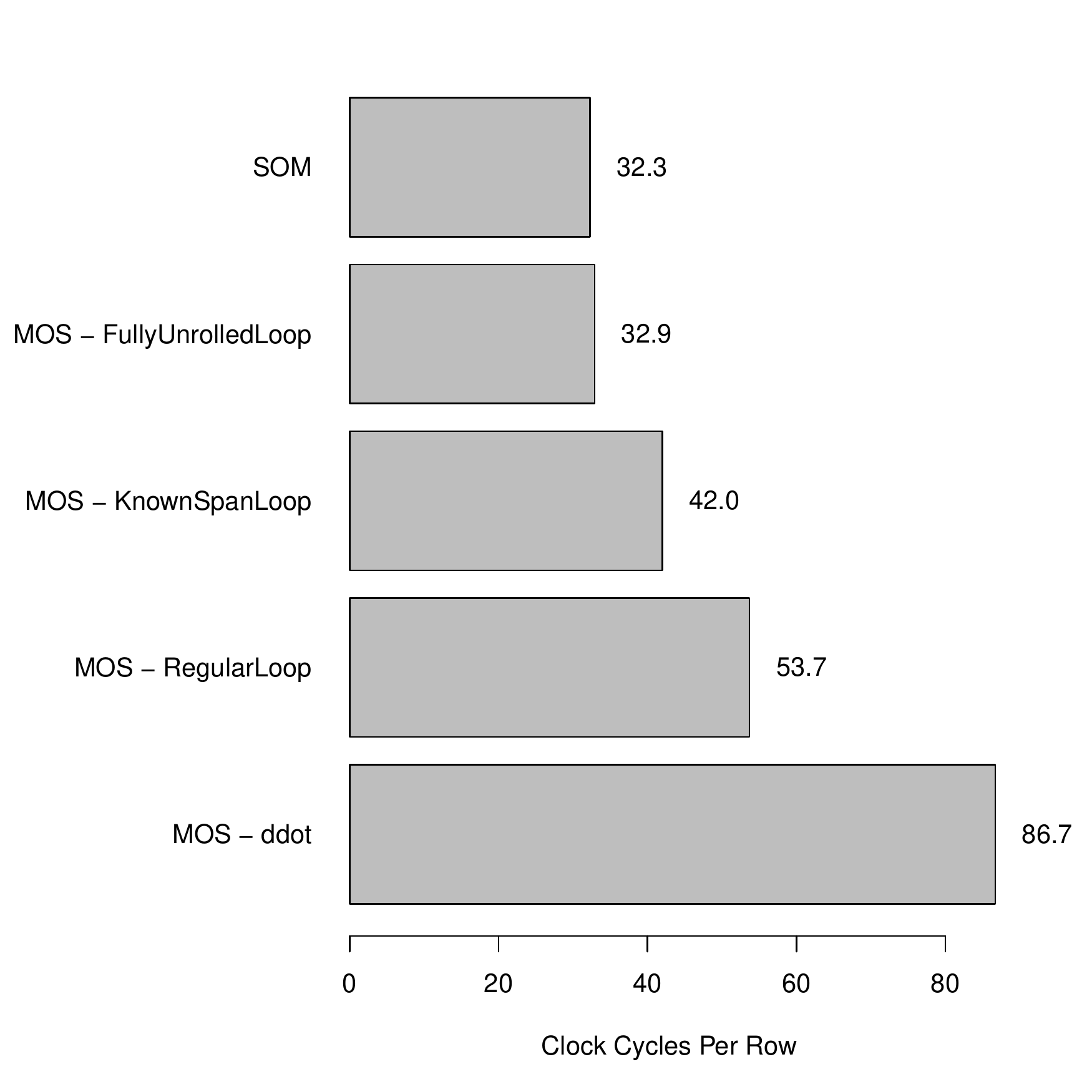}
\caption[]{Performance comparison of SOM and several methods in MOS for implementing the inner product $\xx_n^t \bbeta$. Parameters: $K=10$, $N=1000$. All implementations are vectorized but single-threaded.}
\label{fig:jit-compile}
\end{centering}
\end{figure}

The assembly code for the function in Fig. \ref{fig:code-expanded} (with the fused loop vectorized) consists of three loops, each containing calls to AVX-vectorized transcendental functions \texttt{\_\_svml\_exp4} and \texttt{\_\_svml\_log4}: \footnote{To focus on vectorization, we compiled the code without the OpenMP pragma.} \footnote{For more on generating vectorized code, see this Web Aside \url{http://csapp.cs.cmu.edu/public/waside/waside-simd.pdf} to Bryant and O'Hallaron's great textbook (\cite{bryant2003computer}).} The reason for breaking the loop into 3 parts is to take advantage of efficient data movement instructions for aligned data in the main loop, while allowing for scalar data movement instructions to handle misalignment in the first loop, and array lengths that are incomplete multiples of 4 in the last loop (since each AVX vector instruction handles 4 doubles in its 256-bit registers). The evidence for partial vectorization comes from seeing 4 calls to \texttt{ddot} inside the main loop: Since \texttt{ddot} is scalar (i.e. it produces only one element per call as input to the transcendental functions), it must be called 4 times to generate the full input for the vector instructions. As we see in Fig. \ref{fig:jit-compile}, the resulting performance (`MOS - ddot', CPR of 86.7) is much worse than the SOM pattern of Fig. \ref{fig:code-baseline} (`SOM', CPR of 32.3).

Since \texttt{ddot} is a simple routine, we can replace it with a for loop (`MOS - RegularLoop'), with significant performance improvement. We can further optimize the code by making the loop span a compile-time constant (`MOS - KnownSpanLoop'). Best result, however, is achieved when we fully unroll the loop (`MOS - FullyUnrolledLoop'), producing essentially the same performance as the SOM pattern.

Loop unroll does not necessarily require hand-coding by the programmer. Instead, a source-to-source compilation stage can be utilized, or instead we can utilize the template meta-programming facilities of C++ compilers. Fig. \ref{fig:code-jit-unroll} shows an example code that generates a fully-unrolled version of the \texttt{ddot} BLAS routine for calculating the inner product of two vectors. In the logistic regression example, the only parameter subspace for which we saw a significant advantage from `MOS - FullyUnrolledLoop' over SOM is for short and narrow data and when multi-threading. For example, if we use same parameters as in Fig. \ref{fig:jit-compile} but 4 threads instead of 1, the CPR numbers for these two methods become 12.0 and 19.4, respectively. In addition to being inflexible in handling external function calls, the MOS pattern faces another significant performance challenge,  i.e. excessive synchronization costs in the presence of reduction-to-array operations needed for calculating the derivatives of the log-likelihood function (Section \ref{subsec-deriv}).

\begin{figure}
\begin{verbatim}
--------------------------------------------------------------------------
template<int i>
struct inprod {
  static inline double apply(const double x[], const double y[]) {
    return inprod<i-1>::apply(x,y)+x[i-1]*y[i-1];
  }
};
template<>
struct inprod<1> {
  static inline double apply(const double x[], const double y[]) {
    return x[0]*y[0];
  }
};
inline double ddot_unrolled(double x[], double y[], int K) {
  return inprod<NATT>::apply(x,y);
}
--------------------------------------------------------------------------
\end{verbatim}
\caption[]{Using C++ template meta-programming to implement compile-time loop unroll of the \texttt{ddot} BLAS call, used in the code shown in Fig. \ref{fig:code-expanded}.}
\label{fig:code-jit-unroll}
\end{figure}

Moving many tuning decisions from runtime to compile time is generally a fruitful strategy for HPC. A recent example of this is the Stan compiler for Bayesian DAGs. In contrast to previous compilers such as WINBUGS and JAGS, Stan performs a full compilation of DAG into C++ code. This is likely to be one reason for its superior performance (in addition to using a more efficient sampler in HMC and NUTS). Our manual loop unroll pushes the JIT compiling idea further by requiring the compilation to occur for each new set of values for problem dimensions (specifically $K$ here). This is a minor nuisance but can offer performance gains especially for vectorization.   

\section{Concluding Remarks} \label{sec-discussion}
\subsection{Summary} \label{subsec-discuss-summary}
In this paper, we presented two opportunities for SIMD parallel MCMC sampling of probabilistic graphical models: 1) parallel calculation of log-likelihood and its derivatives for exchangeable models with lack of conjugacy, such as GLM, and 2) parallel sampling of conditionally-independent nodes for graphs with discrete nodes and conjugacy, such as Ising model. In addition, we offered strategies for efficient implementation of such opportunities on multi-core x86 processors using a combination of vectorizing compiler and libraries, and a series of high-level changes to the source code and runtime environment. These strategies included Partial Loop Fusion to minimize parallel overhead, NUMA-aware adjustments to minimize cross-socket communication, cache fusion to maximize cache utilization, and batch RNG for efficient random number generation and code vectorization. For Generalized Linear Models, one of the most commonly-used class of models in predictive analytics, our implementation of the log-likelihood function and its gradient vector reached within 25\% of the hardware limit induced by total memory bandwidth in the big-data regime, outperforming a fully vectorized and (built-in) multi-threaded - but naive - implementation using Intel MKL by more than 5x.

There are two conceptual extremes with regards to utilizing pre-compiled optimized libraries as components of a HPC application: 1) dropping in the pre-compiled library components as they are, and getting suboptimal aggregate results, or 2) discarding all pre-compiled components and writing everything from scratch, which can significantly raise the development costs for the project. Our research suggests a middle ground, through high-level modifications layered on top of pre-compiled libraries, allowing for significant performance improvement while incurring only modest development costs.

\subsection{Future Directions} \label{subsec-discuss-future}
There are several important areas for further research.
\begin{enumerate}
\item \textbf{Portable Implementation of Optimization Techniques:} In order to turn research into action, we must identify robust and portable methods for implementing our proposed optimization techniques. For example, our NUMA adjustment relied on the first-touch policy used by Linux. Also, in cache fusion, the optimal number of chunks is tied to the size of L2 cache. Such hardware and OS dependencies must be addressed before a widely-usable library can be released.
\item \textbf{High-Performance GLM Hessian:} We offered efficient implementation of log-likelihood function and its first derivative for GLM. However, due to their log-concave property [ref], GLMs are suitable for optimization and sampling techniques that utilize the second derivative - Hessian matrix - of the log-likelihood (or log-posterior) function, such as Newton optimization (\cite{nocedal2006numerical}) (equivalently known as Iterative Re-Weighted Least Squares or IRWLS for GLM models) or RM-HMC sampling (\cite{girolami2011riemann}). Developing an equally-efficient implementation of the Hessian for GLMs will be an important contribution to the field of (Bayesian and non-Bayesian) predictive analytics. We expect the Hierachical Loop Structure framework of Section \ref{subsec-deriv} to remain effective, but additional opportunities for optimization, such as taking advantage of the symmetry of the Hessian matrix, must be thoroughly studied.
\item \textbf{Hierarchical Bayesian GLM:} Many Bayesian applications of GLM involves an HB framework. Our research highlights the subtleties of combining coarse- and fine-grained parallel modes available for HB GLMs. These subtleties as well as additional complexities arising from highly-imbalanced regression groups must be carefully studied and encoded into a high-performance library for MCMC sampling of HB GLM models (or generic DAG compilers such as Stan).
\item \textbf{Batch RNG:} We demonstrated the value of batch RNG for improving efficiency and vectorization of MCMC sampling for models such as Ising, RBM, and LDA. For complex distributions, existing libraries cannot be utilized in batch mode in the context of MCMC sampling since the parameters of distributions changes from one iteration to next. (This is in contrast to purely Monte Carlo - non-Markovian - settings where sampled distributions remain stationary throughout the simulation.) Developing batch RNG libraries, i.e. libraries that generate buffers of random deviates for a set of core distributions such as uniform and normal, and utilize these buffers for generating random deviates from complex distributions - will be an important contribution towards high-performance MCMC sampling of big-data simulations such as automated, semantic analysis of web data.
\item \textbf{Vectorization and Atomic Operations:} Atomic operations arise in parallel sampling of conditionally-independent nodes (such as in non-collapsed LDA and in RBM), where nodes may simultaneously attempt to update the same intermediate data structures. Before hardware support for vectorized atomic operations become available, software options can be explored. For example, one possibility is to accumulate updates in temporary buffers such that each vector instruction produces non-overlapping updates to the buffer. Higher flexibility offered by vector intrinsics might be required to implement such solutions.
\end{enumerate}

\parindent0pt \textbf{References}

\bibliographystyle{elsarticle-harv} 
\bibliography{pargibbs}

\begin{thebibliography}{71}
\expandafter\ifx\csname natexlab\endcsname\relax\def\natexlab#1{#1}\fi
\expandafter\ifx\csname url\endcsname\relax
  \def\url#1{\texttt{#1}}\fi
\expandafter\ifx\csname urlprefix\endcsname\relax\def\urlprefix{URL }\fi

\bibitem[{Ackley et~al.(1985)Ackley, Hinton, and
  Sejnowski}]{ackley1985learning}
Ackley, D.~H., Hinton, G.~E., Sejnowski, T.~J., 1985. A learning algorithm for
  boltzmann machines*. Cognitive science 9~(1), 147--169.

\bibitem[{Ahmed et~al.(2012)Ahmed, Aly, Gonzalez, Narayanamurthy, and
  Smola}]{ahmed2012scalable}
Ahmed, A., Aly, M., Gonzalez, J., Narayanamurthy, S., Smola, A.~J., 2012.
  Scalable inference in latent variable models. In: Proceedings of the fifth
  ACM international conference on Web search and data mining. ACM, pp.
  123--132.

\bibitem[{Antonio and Beirlant(2007)}]{antonio2007actuarial}
Antonio, K., Beirlant, J., 2007. Actuarial statistics with generalized linear
  mixed models. Insurance: Mathematics and Economics 40~(1), 58--76.

\bibitem[{Babapulle et~al.(2004)Babapulle, Joseph, B{\'e}lisle, Brophy, and
  Eisenberg}]{babapulle2004hierarchical}
Babapulle, M.~N., Joseph, L., B{\'e}lisle, P., Brophy, J.~M., Eisenberg, M.~J.,
  2004. A hierarchical bayesian meta-analysis of randomised clinical trials of
  drug-eluting stents. The Lancet 364~(9434), 583--591.

\bibitem[{Bache and Lichman(2013)}]{BacheLichman2013}
Bache, K., Lichman, M., 2013. {UCI} machine learning repository.
\newline\urlprefix\url{http://archive.ics.uci.edu/ml}

\bibitem[{Bernardo(1996)}]{bernardo1996concept}
Bernardo, J.~M., 1996. The concept of exchangeability and its applications. Far
  East Journal of Mathematical Sciences 4, 111--122.

\bibitem[{Bishop(2006)}]{christopher2006pattern}
Bishop, C.~M., 2006. Pattern recognition and machine learning.

\bibitem[{Blei(2012)}]{blei2012probabilistic}
Blei, D.~M., 2012. Probabilistic topic models. Communications of the ACM
  55~(4), 77--84.

\bibitem[{Bovet and Cesati(2005)}]{bovet2005understanding}
Bovet, D.~P., Cesati, M., 2005. Understanding the Linux kernel. O'Reilly Media,
  Inc.

\bibitem[{Brockwell(2006)}]{brockwell2006parallel}
Brockwell, A., 2006. Parallel markov chain monte carlo simulation by
  pre-fetching. Journal of Computational and Graphical Statistics 15~(1),
  246--261.

\bibitem[{Broquedis et~al.(2009)Broquedis, Furmento, Goglin, Namyst, and
  Wacrenier}]{broquedis2009dynamic}
Broquedis, F., Furmento, N., Goglin, B., Namyst, R., Wacrenier, P.-A., 2009.
  Dynamic task and data placement over numa architectures: an openmp runtime
  perspective. In: Evolving OpenMP in an Age of Extreme Parallelism. Springer,
  pp. 79--92.

\bibitem[{Brush(1967)}]{brush1967history}
Brush, S.~G., 1967. History of the lenz-ising model. Reviews of Modern Physics
  39~(4), 883.

\bibitem[{Bryant and O'Hallaron(2011)}]{bryant2003computer}
Bryant, R., O'Hallaron, D.~R., 2011. Computer systems: a programmer's
  perspective, Second Edition. Prentice Hall.

\bibitem[{Carlin et~al.(1992)Carlin, Gelfand, and
  Smith}]{carlin1992hierarchical}
Carlin, B.~P., Gelfand, A.~E., Smith, A.~F., 1992. Hierarchical bayesian
  analysis of changepoint problems. Applied statistics, 389--405.

\bibitem[{Chandra(2001)}]{chandra2001parallel}
Chandra, R., 2001. Parallel programming in OpenMP. Morgan Kaufmann.

\bibitem[{da~Silva(2011)}]{da2011cudabayesreg}
da~Silva, A. R.~F., 2011. Cudabayesreg: parallel implementation of a bayesian
  multilevel model for fmri data analysis. Journal of Statistical Software
  44~(4), 1--24.

\bibitem[{Descombes et~al.(1998)Descombes, Kruggel, and von
  Cramon}]{descombes1998spatio}
Descombes, X., Kruggel, F., von Cramon, D.~Y., 1998. Spatio-temporal fmri
  analysis using markov random fields. Medical Imaging, IEEE Transactions on
  17~(6), 1028--1039.

\bibitem[{Diaconescu and Zima(2007)}]{diaconescu2007approach}
Diaconescu, R.~E., Zima, H.~P., 2007. An approach to data distributions in
  chapel. International Journal of High Performance Computing Applications
  21~(3), 313--335.

\bibitem[{Duane et~al.(1987)Duane, Kennedy, Pendleton, and
  Roweth}]{duane1987hybrid}
Duane, S., Kennedy, A.~D., Pendleton, B.~J., Roweth, D., 1987. Hybrid monte
  carlo. Physics letters B 195~(2), 216--222.

\bibitem[{Dumont(2011)}]{dumont2011markov}
Dumont, M.~D., 2011. Markov chain monte carlo on the gpu. Ph.D. thesis,
  Rochester Institute of Technology.

\bibitem[{Fenton and Neil(2004)}]{fenton2004combining}
Fenton, N., Neil, M., 2004. Combining evidence in risk analysis using bayesian
  networks. Agena White Paper W 704.

\bibitem[{Gelman and Hill(2007)}]{gelman2007data}
Gelman, A., Hill, J., 2007. Data analysis using regression and
  multilevel/hierarchical models. Cambridge University Press.

\bibitem[{Geman and Geman(1984)}]{geman1984stochastic}
Geman, S., Geman, D., 1984. Stochastic relaxation, gibbs distributions, and the
  bayesian restoration of images. IEEE Transactions on Pattern Analysis and
  Machine Intelligence~(6), 721--741.

\bibitem[{Gilks and Wild(1992)}]{gilks1992adaptive}
Gilks, W.~R., Wild, P., 1992. Adaptive rejection sampling for gibbs sampling.
  Applied Statistics, 337--348.

\bibitem[{Girolami and Calderhead(2011)}]{girolami2011riemann}
Girolami, M., Calderhead, B., 2011. Riemann manifold langevin and hamiltonian
  monte carlo methods. Journal of the Royal Statistical Society: Series B
  (Statistical Methodology) 73~(2), 123--214.

\bibitem[{Gonzalez et~al.(2011)Gonzalez, Low, Gretton, and
  Guestrin}]{gonzalez2011parallel}
Gonzalez, J., Low, Y., Gretton, A., Guestrin, C., 2011. Parallel gibbs
  sampling: From colored fields to thin junction trees. In: International
  Conference on Artificial Intelligence and Statistics. pp. 324--332.

\bibitem[{Gonzalez et~al.(2012)Gonzalez, Low, Gu, Bickson, and
  Guestrin}]{gonzalez2012powergraph}
Gonzalez, J.~E., Low, Y., Gu, H., Bickson, D., Guestrin, C., 2012. Powergraph:
  Distributed graph-parallel computation on natural graphs. In: Proceedings of
  the 10th USENIX Symposium on Operating Systems Design and Implementation
  (OSDI). pp. 17--30.

\bibitem[{Gonzalez et~al.(2009)Gonzalez, Low, Guestrin, and
  O'Hallaron}]{gonzalez2009distributed}
Gonzalez, J.~E., Low, Y., Guestrin, C., O'Hallaron, D., 2009. Distributed
  parallel inference on large factor graphs. In: Proceedings of the
  Twenty-Fifth Conference on Uncertainty in Artificial Intelligence. AUAI
  Press, pp. 203--212.

\bibitem[{Hastie and Tibshirani(1987)}]{hastie1987non}
Hastie, T., Tibshirani, R., 1987. Non-parametric logistic and proportional odds
  regression. Applied statistics 36~(3), 260--276.

\bibitem[{Hastings(1970)}]{hastings1970monte}
Hastings, W.~K., 1970. Monte carlo sampling methods using markov chains and
  their applications. Biometrika 57~(1), 97--109.

\bibitem[{Hinton et~al.(2006)Hinton, Osindero, and Teh}]{hinton2006fast}
Hinton, G.~E., Osindero, S., Teh, Y.-W., 2006. A fast learning algorithm for
  deep belief nets. Neural computation 18~(7), 1527--1554.

\bibitem[{Hoffman and Gelman(2011)}]{hoffman2011no}
Hoffman, M.~D., Gelman, A., 2011. The no-u-turn sampler: Adaptively setting
  path lengths in hamiltonian monte carlo. arXiv preprint arXiv:1111.4246.

\bibitem[{Hopfield(1982)}]{hopfield1982neural}
Hopfield, J.~J., 1982. Neural networks and physical systems with emergent
  collective computational abilities. Proceedings of the national academy of
  sciences 79~(8), 2554--2558.

\bibitem[{Jeffers and Reinders(2013)}]{jeffers2013intel}
Jeffers, J., Reinders, J., 2013. Intel Xeon Phi Coprocessor High Performance
  Programming. Newnes.

\bibitem[{Kirk and Wen-mei(2012)}]{kirk2012programming}
Kirk, D.~B., Wen-mei, W.~H., 2012. Programming massively parallel processors: a
  hands-on approach. Newnes.

\bibitem[{Krizhevsky et~al.(2012)Krizhevsky, Sutskever, and
  Hinton}]{krizhevsky2012imagenet}
Krizhevsky, A., Sutskever, I., Hinton, G.~E., 2012. Imagenet classification
  with deep convolutional neural networks. In: NIPS. Vol.~1. p.~4.

\bibitem[{Kumar et~al.(2008)Kumar, Kim, Smelyanskiy, Chen, Chhugani, Hughes,
  Kim, Lee, and Nguyen}]{kumar2008atomic}
Kumar, S., Kim, D., Smelyanskiy, M., Chen, Y.-K., Chhugani, J., Hughes, C.~J.,
  Kim, C., Lee, V.~W., Nguyen, A.~D., 2008. Atomic vector operations on chip
  multiprocessors. In: ACM SIGARCH Computer Architecture News. Vol.~36. IEEE
  Computer Society, pp. 441--452.

\bibitem[{Le et~al.(2011)Le, Zou, Yeung, and Ng}]{le2011learning}
Le, Q.~V., Zou, W.~Y., Yeung, S.~Y., Ng, A.~Y., 2011. Learning hierarchical
  invariant spatio-temporal features for action recognition with independent
  subspace analysis. In: Computer Vision and Pattern Recognition (CVPR), 2011
  IEEE Conference on. IEEE, pp. 3361--3368.

\bibitem[{Lee and Mumford(2003)}]{lee2003hierarchical}
Lee, T.~S., Mumford, D., 2003. Hierarchical bayesian inference in the visual
  cortex. JOSA A 20~(7), 1434--1448.

\bibitem[{Low et~al.(2012)Low, Bickson, Gonzalez, Guestrin, Kyrola, and
  Hellerstein}]{low2012distributed}
Low, Y., Bickson, D., Gonzalez, J., Guestrin, C., Kyrola, A., Hellerstein,
  J.~M., 2012. Distributed graphlab: a framework for machine learning and data
  mining in the cloud. Proceedings of the VLDB Endowment 5~(8), 716--727.

\bibitem[{Majo and Gross(2011)}]{majo2011memory}
Majo, Z., Gross, T.~R., 2011. Memory management in numa multicore systems:
  trapped between cache contention and interconnect overhead. In: ACM SIGPLAN
  Notices. Vol.~46. ACM, pp. 11--20.

\bibitem[{Marsaglia and Tsang(2000)}]{marsaglia2000simple}
Marsaglia, G., Tsang, W.~W., 2000. A simple method for generating gamma
  variables. ACM Transactions on Mathematical Software (TOMS) 26~(3), 363--372.

\bibitem[{McCool et~al.(2012)McCool, Reinders, and
  Robison}]{mccool2012structured}
McCool, M., Reinders, J., Robison, A., 2012. Structured parallel programming:
  patterns for efficient computation. Elsevier.

\bibitem[{McCulloch(2006)}]{mcculloch2006generalized}
McCulloch, C.~E., 2006. Generalized linear mixed models. Wiley Online Library.

\bibitem[{Metropolis et~al.(2004)Metropolis, Rosenbluth, Rosenbluth, Teller,
  and Teller}]{metropolis2004equation}
Metropolis, N., Rosenbluth, A.~W., Rosenbluth, M.~N., Teller, A.~H., Teller,
  E., 2004. Equation of state calculations by fast computing machines. The
  journal of chemical physics 21~(6), 1087--1092.

\bibitem[{Metropolis and Ulam(1949)}]{metropolis1949monte}
Metropolis, N., Ulam, S., 1949. The monte carlo method. Journal of the American
  statistical association 44~(247), 335--341.

\bibitem[{Mohamed et~al.(2011)Mohamed, Sainath, Dahl, Ramabhadran, Hinton, and
  Picheny}]{mohamed2011deep}
Mohamed, A.-r., Sainath, T.~N., Dahl, G., Ramabhadran, B., Hinton, G.~E.,
  Picheny, M.~A., 2011. Deep belief networks using discriminative features for
  phone recognition. In: Acoustics, Speech and Signal Processing (ICASSP), 2011
  IEEE International Conference on. IEEE, pp. 5060--5063.

\bibitem[{Molina~da Cruz et~al.(2011)Molina~da Cruz, Alves, Carissimi, Navaux,
  Ribeiro, and Mehaut}]{molina2011using}
Molina~da Cruz, E.~H., Alves, Z., Carissimi, A., Navaux, P. O.~A., Ribeiro,
  C.~P., Mehaut, J., 2011. Using memory access traces to map threads and data
  on hierarchical multi-core platforms. In: Parallel and Distributed Processing
  Workshops and Phd Forum (IPDPSW), 2011 IEEE International Symposium on. IEEE,
  pp. 551--558.

\bibitem[{Neal(1995)}]{neal1995bayesian}
Neal, R.~M., 1995. Bayesian learning for neural networks. Ph.D. thesis,
  University of Toronto.

\bibitem[{Neal(2003)}]{neal2003slice}
Neal, R.~M., 2003. Slice sampling. Annals of statistics, 705--741.

\bibitem[{Nelder and Baker(1972)}]{nelder1972generalized}
Nelder, J.~A., Baker, R., 1972. Generalized linear models. Wiley Online
  Library.

\bibitem[{Newman et~al.(2009)Newman, Asuncion, Smyth, and
  Welling}]{newman2009distributed}
Newman, D., Asuncion, A., Smyth, P., Welling, M., 2009. Distributed algorithms
  for topic models. The Journal of Machine Learning Research 10, 1801--1828.

\bibitem[{Nocedal and Wright(2006)}]{nocedal2006numerical}
Nocedal, J., Wright, S., 2006. Numerical optimization: Springer series in
  operations research and financial engineering. Springer-Verlag.

\bibitem[{Perez et~al.(1998)}]{perez1998markov}
Perez, P., et~al., 1998. Markov random fields and images. CWI quarterly 11~(4),
  413--437.

\bibitem[{Polikar(2006)}]{polikar2006ensemble}
Polikar, R., 2006. Ensemble based systems in decision making. Circuits and
  Systems Magazine, IEEE 6~(3), 21--45.

\bibitem[{Press(2007)}]{press2007numerical}
Press, W.~H., 2007. Numerical recipes 3rd edition: The art of scientific
  computing. Cambridge university press.

\bibitem[{Qi and Minka(2002)}]{qi2002hessian}
Qi, Y., Minka, T.~P., 2002. Hessian-based markov chain monte-carlo algorithms.

\bibitem[{Ren and Orkoulas(2007)}]{ren2007parallel}
Ren, R., Orkoulas, G., 2007. Parallel markov chain monte carlo simulations. The
  Journal of chemical physics 126~(21), 211102.

\bibitem[{Ribeiro et~al.(2010)Ribeiro, M{\'e}haut, and
  Carissimi}]{ribeiro2010memory}
Ribeiro, C.~P., M{\'e}haut, J., Carissimi, A., 2010. Memory affinity management
  for numerical scientific applications over multi-core multiprocessors with
  hierarchical memory. In: Parallel \& Distributed Processing, Workshops and
  Phd Forum (IPDPSW), 2010 IEEE International Symposium on. IEEE, pp. 1--4.

\bibitem[{Ripley(2009)}]{ripley2009stochastic}
Ripley, B.~D., 2009. Stochastic simulation. Vol. 316. John Wiley \& Sons.

\bibitem[{Robert and Casella(2004)}]{robert2004monte}
Robert, C.~P., Casella, G., 2004. Monte Carlo statistical methods. Vol. 319.
  Citeseer.

\bibitem[{Rossi et~al.(2005)Rossi, Allenby, and McCulloch}]{peter2005bayesian}
Rossi, P.~E., Allenby, G.~M., McCulloch, R.~E., 2005. Bayesian statistics and
  marketing. J. Wiley \& Sons.

\bibitem[{Sharabiani et~al.(2011)Sharabiani, Vermeulen, Scoccianti, Hosnijeh,
  Minelli, Sacerdote, Palli, Krogh, Tumino, Chiodini,
  et~al.}]{azar2011immunologic}
Sharabiani, M. T.~A., Vermeulen, R., Scoccianti, C., Hosnijeh, F.~S., Minelli,
  L., Sacerdote, C., Palli, D., Krogh, V., Tumino, R., Chiodini, P., et~al.,
  2011. Immunologic profile of excessive body weight. Biomarkers 16~(3),
  243--251.

\bibitem[{Smolensky(1986)}]{smolensky1986information}
Smolensky, P., 1986. Information processing in dynamical systems: Foundations
  of harmony theory.

\bibitem[{Sobehart et~al.(2000)Sobehart, Stein, Mikityanskaya, and
  Li}]{sobehart2000moody}
Sobehart, J.~R., Stein, R., Mikityanskaya, V., Li, L., 2000. Moody’s public
  firm risk model: A hybrid approach to modeling short term default risk.
  Moody’s Investors Service, Global Credit Research, Rating Methodology,
  March.

\bibitem[{Strid(2010)}]{strid2010efficient}
Strid, I., 2010. Efficient parallelisation of metropolis--hastings algorithms
  using a prefetching approach. Computational Statistics \& Data Analysis
  54~(11), 2814--2835.

\bibitem[{Sutter and Kalivas(1993)}]{sutter1993comparison}
Sutter, J.~M., Kalivas, J.~H., 1993. Comparison of forward selection, backward
  elimination, and generalized simulated annealing for variable selection.
  Microchemical journal 47~(1), 60--66.

\bibitem[{Tibbits et~al.(2011)Tibbits, Haran, and
  Liechty}]{tibbits2011parallel}
Tibbits, M.~M., Haran, M., Liechty, J.~C., 2011. Parallel multivariate slice
  sampling. Statistics and Computing 21~(3), 415--430.

\bibitem[{Wilkinson(2010)}]{wilkinson2010parallel}
Wilkinson, D.~J., 2010. Parallel bayesian computation. In: Kontoghiorghes,
  E.~J. (Ed.), Handbook of Parallel Computing and Statistics. CRC Press.

\bibitem[{Xu and Ihler(2011)}]{xu2011multicore}
Xu, T., Ihler, A.~T., 2011. Multicore gibbs sampling in dense, unstructured
  graphs. In: International Conference on Artificial Intelligence and
  Statistics. pp. 798--806.

\bibitem[{Yu and Xu(2009)}]{yu2009parallel}
Yu, L., Xu, Y., 2009. A parallel gibbs sampling algorithm for motif finding on
  gpu. In: Parallel and Distributed Processing with Applications, 2009 IEEE
  International Symposium on. IEEE, pp. 555--558.

\end{thebibliography}





\end{document}